\documentclass[aps,prb,reprint,superscriptaddress]{revtex4-1}
\usepackage{epsf}
\usepackage{graphicx}
\usepackage{gensymb}
\usepackage{natbib}
\usepackage{float}
\usepackage{appendix}
\usepackage{gensymb}
\usepackage{upgreek}

\newcommand{\overbar}[1]{\mkern 1.5mu\overline{\mkern-1.5mu#1\mkern-1.5mu}\mkern 1.5mu}

\begin{document}
\title{A sharp increase in the density of states in PbTe approaching a saddle point in the band structure.}

\author{P. Walmsley}
\affiliation{Department of Applied Physics and Geballe Laboratory
for Advanced Materials, Stanford University, Stanford, California
94305, USA}

\author{D. M. Abrams}
\affiliation{Department of Applied Physics and Geballe Laboratory
for Advanced Materials, Stanford University, Stanford, California
94305, USA}

\author{J. Straquadine}
\affiliation{Department of Applied Physics and Geballe Laboratory
for Advanced Materials, Stanford University, Stanford, California
94305, USA}

\author{M. K. Chan}
\affiliation{Los Alamos National Laboratory, Los Alamos, NM 87545, USA}

\author{R. D. McDonald}
\affiliation{Los Alamos National Laboratory, Los Alamos, NM 87545, USA}

\author{P. Giraldo-Gallo}
\affiliation{Department of Physics, Universidad de Los Andes, Bogot\'{a} 111711, Colombia}


\author{I. R. Fisher}
\affiliation{Department of Applied Physics and Geballe Laboratory
for Advanced Materials, Stanford University, Stanford, California
94305, USA}

\date{\today}
\begin{abstract}
PbTe is a leading mid-range thermoelectric material with a $zT$ that has been enhanced by, amongst other methods, band engineering. Here we present an experimental study of the Hall effect, quantum oscillations, specific heat, and electron microprobe analysis that explores the evolution of the electronic structure of PbTe heavily doped with the `ideal' acceptor Na up to the solubility limit. We identify two phenomenological changes that onset as the electronic structure deviates from a Kane-type dispersion at around 180\,meV; a qualitative change in the field dependence of the Hall effect indicative of an increase in the high-field limit and a change in the Fermiology, and a sharp increase in the density of states as a function of energy. Following consideration of three possible origins for the observed phenomenology we conclude that the most likely source is non-ellipsoidicity of the $L$-pocket upon approach to a saddle point in the band structure, which is evidenced directly by our quantum oscillation measurements. Comparison to density functional theory calculations imply that this evolution of the electronic structure may be a key contributor to the large thermopower in PbTe.
\end{abstract}

\maketitle

\section{Introduction}

PbTe is a narrow band gap semiconductor that has provoked significant interest in both applications and fundamental science for several decades. It has recently received renewed attention in the context of thermoelectric power generation owing to a reported figure of merit, $zT$, greater than 2 acheived upon optimisation of the band structure and phonon scattering\cite{Pei2012, Biswas2012,Wu2014, Zhao2013}, and $zT$ of up 1.5 via the introduction of resonant impurity states\cite{Heremans2008}. From the perspective of fundamental science, PbTe exhibits a number of properties that are of current interest including strong phonon anharmonicity at low energies\cite{Delaire2011}, proximity to a topological phase transition\cite{Xu2012, Dziawa2012}, anomalously high-temperature superconductivity\cite{Chernik1981, Matsushita2006, Giraldo-Gallo2017}, and evidence for a charge-Kondo effect\cite{Matsushita2005}. Central to understanding of these phenomena, and the ability to tune them for applications, is a detailed understanding of the electronic structure.

Both experimental investigations and calculations of the electronic structure in PbTe show some significant disagreement on the relative energies at which the valence band should deviate from a single-band Kane-type dispersion, whose key features are a constant anisotropy (ellipsoidicity) and a linear dependence of the electronic effective mass as a function of energy close to the band extrema. It also remains unclear whether the band structure generally deviates from Kane-type via the appearance of a second valence-band maximum, significant non-ellipsoidicity, or resonant impurity states at the Fermi level in PbTe\cite{Singh2010, Giraldo-Gallo2016, Ahmad2006, Xiong2010, Ravich1970, Nimtz1983, Allgaier1966, Bilc2006}. As each of these scenarios should increase the density of states, the Mott relation dictates that the distinction between these scenarios becomes important in engineering a high $zT$. The situation is further complicated by the temperature dependence of both the direct gap and the offset of the first and second valence band maxima (which are thought to converge at elevated temperatures)\cite{Gibbs2013, Zhao2014, Ravich1970, Nimtz1983}, and the sensitivity of the band edges to perturbations such as pressure, temperature, chemical composition and spin-orbit coupling \cite{Ravich1970, Nimtz1983}. While extensive experimental studies of the thermal and electrical properties of PbTe at elevated temperatures have indirectly inferred information about its band structure, there has been comparatively few studies directly characterising the electronic structure away from the band edge and in the groundstate.

\begin{figure}
\includegraphics[width=3.2in]{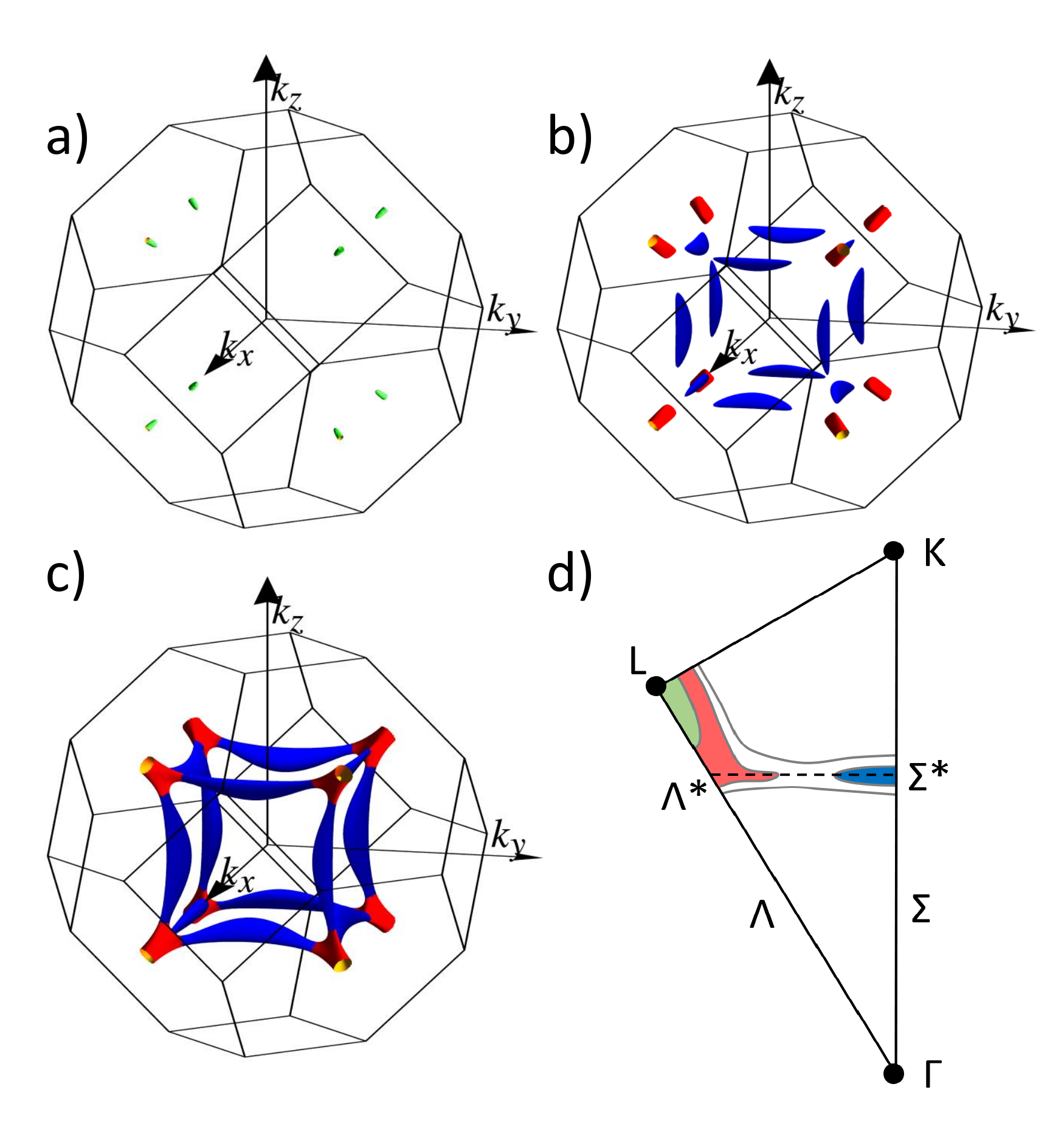}
\caption{a)-c) Calculations of the Fermi surface of PbTe at successively larger Fermi energies, presented previously by Giraldo-Gallo \emph{et al.}\cite{Giraldo-Gallo2016} a) is representative of the single-band ellipsoid found at the $L$-point at low to intermediate dopings, b) shows the second band maximum (blue) as well as non-ellipsoidicity in the $L$-pocket (red), and c) shows the anticipated Fermi surface upon crossing the saddle point connecting the two upper valence band maxima. d) is an illustration of constant energy surfaces in the $L-K-\Gamma$ plane exaggerated for clarity. Green, red and blue respectively illustrate an ellipsoidal $L$-pocket, the anticipated non-ellipsoidicity of the $L$-pocket and the $\Sigma$ pocket, as in a) and b). We define $\Sigma^*$ as the position of the second valence band maximum, which is expected to connect to the $L$-pocket along the $\Sigma^*-\Lambda^*$ line at large values of the Fermi energy.}
\label{FIG:BSsketch}
\end{figure}

Recently, a detailed quantum oscillation study by Giraldo-Gallo \emph{et al.}\cite{Giraldo-Gallo2016} extended previous works\cite{Jensen1978} to characterise the Fermi surface in Na-doped PbTe (Pb$_{1-x}$Na$_x$Te) down to 160\,meV below the valence band edge. It was shown that the Fermi surface is fully described by ellipsoid of fixed ellipsoidicity located at the $L$-point with its  major semi-axis along the $\Lambda$ line, as shown in Figure \ref{FIG:BSsketch}a and illustrated in green in Figure \ref{FIG:BSsketch}d. The results were consistent with a Kane-type dispersion at all hole concentrations studied ($p < $9$\times$10$^{19}$\,cm$^{-3}$) \cite{Giraldo-Gallo2016, Giraldo-Gallo2017}. It has been known for some time from elevated temperature studies of hot electrons that the second band maximum is anisotropic, forming an elongated pocket oriented along the [100] direction and occuring along the $\Sigma$ line (at a point referred to here as $\Sigma^*$ and shown in blue in Figures \ref{FIG:BSsketch}b and \ref{FIG:BSsketch}d)\cite{Sitter1977, Nimtz1983}, and this has been confirmed more recently by ARPES studies that probe states below the Fermi level, $E_F$\cite{Zhao2014, Nakayama2008, Giraldo-Gallo2017}. These two band maxima are connected by a saddle point in the band structure along the line connecting the points $\Sigma^*$ and $\Lambda^*$ as defined in Figure \ref{FIG:BSsketch}d that is expected to join the two band maxima to form a single cage-like Fermi surface (Figure \ref{FIG:BSsketch}c) at an energy not far below that of the second valence band maximum\cite{Singh2010, Giraldo-Gallo2016}. As such, at some energy the $L$-pocket must become non-ellipsoidal in a manner highlighted in red in Figures \ref{FIG:BSsketch}b and \ref{FIG:BSsketch}d, deviating from the Kane model as it does so and enhancing the density of states. It is unclear from existing experiments in what order these deviations from the single-band Kane model occur and how influential each may be in enhancing the density of states and thus also the thermopower.

In this work we focus on heavily doped samples up to the highest reported Hall numbers acheived by Na doping in PbTe. Data is presented from a range of complimentary techniques (quantum oscillations, specific heat, Hall effect and electron microprobe analysis (EMPA)) from which it is robustly established that the band structure is no longer well described by a single-band Kane-type dispersion below 180\,meV. The most striking feature of this deviation and the key result of this work is that the density of states increases sharply as a function of the Fermi energy. This occurs concurrently with an increase in the high-field threshold observed in the Hall effect which shows that the carrier density is in fact considerably lower than that estimated at generally accessible magnetic fields. This demonstrates that the use of the Hall effect in determining the carrier density and composition of doped PbTe is flawed at high dopings. Consistently with this increase in the high-field threshold, the quantum oscillation measurements cannot resolve the whole Fermi surface, but do seem to show a deviation from the fixed-ellipsoidicity found closer to the band edge. We discuss the relative merits of the  electronic structures proposed above, and conclude that our data favours non-ellipsoidicity of the $L$-pocket on approach to the saddle point in the band structure as the most likely source of the deviation from a Kane-type dispersion.

\section{Methods}
Single crystals were grown via a self-flux technique in order to match the highest reported Hall numbers (a physical vapour transport technique was used in our previous studies\cite{Giraldo-Gallo2016, Matsushita2006}). Crystals were grown in a Te-rich melt to avoid counter-doping by Te vacancies using a Te:(Pb+Na) molar ratio of 70:30. Alumina crucibles with a strainer component were used (ACP-CCS, LSP ceramics). The melt was held at 900$^{\circ}$C and then slow cooled to 550$^{\circ}$C over 3-4 days, with the flux then separated from the crystals by quenching in a centrifuge. The approximate ratio of nominal to actual dopant concentration was found to be around 10:1 below the solubility limit. The resultant single crystals varied in size as doping increased from several mm at low dopings down to around 500\,$\upmu$m to a side at the highest sodium concentrations \cite{Yamini2013}.

Quantum oscillations were measured in two crystals using a mutual inductance technique\cite{VanDegrift1975} up to 65\,T at the National High Magnetic Field Laboratory (NHMFL) in Los Alamos. High-field Hall effect measurements up to 30\,T were performed at the NHMFL in Tallahassee, with the other Hall effect and quantum oscillation measurements taken in a commercially available 14\,T PPMS from Quantum Design and a 16\,T system from Cryogenic Ltd. A standard transverse contact geometry was used with the Hall component isolated by symmetrising between positive and negative fields applied in the [100] direction.  Specific heat measurements were taken on a mosaic of single crystals sourced from the same batch in a commercially available dilution refrigerator option for the PPMS from Quantum Design. The lower temperature range was necessitated by the small values of the Sommerfeld coefficient and higher-order phonon terms observed to low temperature in PbTe which make extrapolations from He$^4$ temperatures unreliable. EMPA was performed on the very same two crystals used in the quantum oscillation measurements (as well as two well characterised, physical vapour transport grown samples from our previous study\cite{Giraldo-Gallo2016}) and found the crystals to be homogenous and single phase (see Appendix \ref{APP:EMPA}), ruling out the presence of precipitates that have been found previously in quenched solid solutions\cite{Yamini2013}. EMPA was performed with a JEOL JXA-8230 SuperProbe Electron Probe Microanalyzer at the Stanford Microchemical Analysis Facility. 

\section{Results}
\subsection{Hall Effect}
\begin{figure}[H]
\includegraphics[width=\columnwidth]{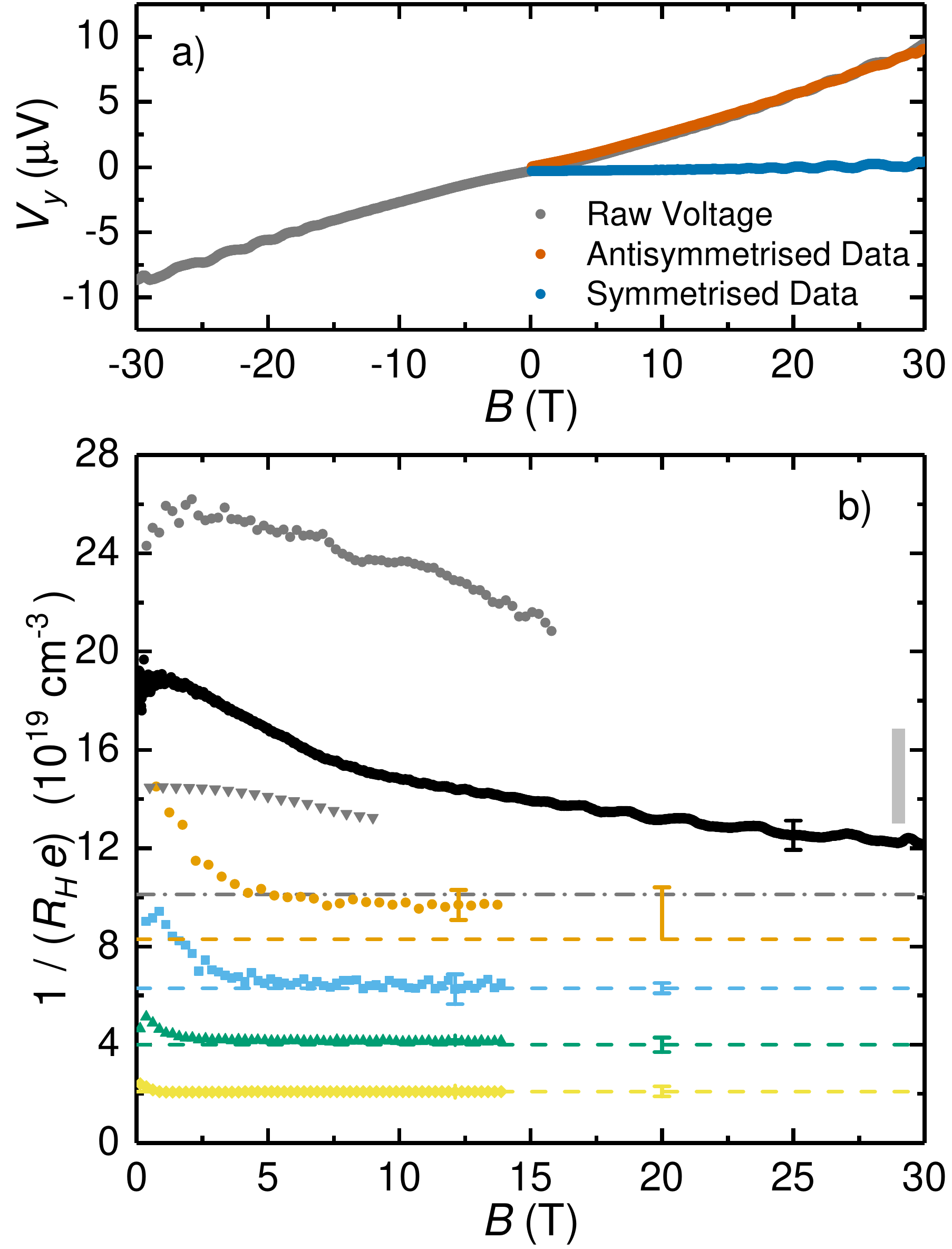}
\caption{a) Raw, symmetrised, and antisymmetrised transverse voltage data, $V_{y}(B)$ (grey, blue, and red lines respectively) up to 30\,T at 1.55\,K. b) The Hall number as a function of field compared to the measured Luttinger volume for samples of varying carrier concentration, measured in each case at a fixed temperature between 1.5\,K and 4\,K. The yellow, green, blue and orange points show the Hall number for samples previously presented by Giraldo-Gallo \emph{et al.}\cite{Giraldo-Gallo2016}, with dashed lines of matching colour showing the corresponding measured Luttinger volumes from the same work.  The black points are derived from the data in a), and the grey points taken on the same two heavily doped samples for which quantum oscillation measurements were performed in pulsed fields up to 65\,T and presented below. The grey dot-dashed line shows a lower-bound for the $L$-pocket Luttinger volume in this sample derived from the present quantum oscillation measurements (as described in the main text). The grey bar to the right shows the density of Na dopants measured directly by EMPA on the same sample as both the top curve in b) (grey circles) and the quantum oscillation measurements presented below (see Appendix \ref{APP:EMPA}), with the dopant density expected to approximately match $p$ in this case.} \label{FIG:HallData}
\end{figure}

In order to clarify the categorisation of heavily doped samples we begin by discussing measurements of the Hall effect. In the high-field limit ($\omega_c\tau \gg 1$) the Hall number is a good measure of the net carrier density, but at lower fields it is acutely sensitive to anisotropy and the presence of multiple inequivalent Fermi-surface pockets.  Figure \ref{FIG:HallData}a shows raw, symmetrised and anti-symmetrised $V_{y}$ data in magnetic fields up to 30\,T for a heavily doped sample of Pb$_{1-x}$Na$_x$Te, with the component that is antisymmetric in $B$ representing the Hall effect. In Figure \ref{FIG:HallData}b this data is replotted as the Hall number ($p_H=1/R_H e$) and compared to samples of lower doping that were characterised in the previous study by Giraldo-Gallo \emph{et al.}\cite{Giraldo-Gallo2016}. At lower dopings, the high field limit is clearly reached at accessible fields as the Hall number becomes a constant that matches the Luttinger volume as measured by quantum oscillations (shown by the dashed lines) in the same study. The Hall number in the present heavily doped sample (black points) however does not reach a constant value even at 30\,T, implying that $\omega_c\tau < 1$ and therefore the Hall number is not a good measure of the carrier density in this doping range. With this in mind it is important that we make a clear distinction between the real hole density in our samples, $p$, and the low-field Hall number, $p_H(B\to 0)$, that is used as a practical means to differentiate our samples and compare to published values. Furthermore, the qualitative evolution of the Hall number with field appears different in this regime which may be indicative of a change in the electronic structure. The grey points were taken from the same samples studied here by quantum oscillations, one of which (grey circles) appears to be at the solubility limit as the data matches the highest reported low-temperature Hall numbers for Pb$_{1-x}$Na$_x$Te\cite{Crocker1967, Yamini2013}. The grey bar in Figure \ref{FIG:HallData}b shows the Na density in the same sample as measured by EMPA, which is expected to match $p$ as Na is an ideal monovalent hole dopant in PbTe (see Appendix \ref{APP:EMPA}).

\subsection{Quantum Oscillations}
\begin{figure*}
\includegraphics[width=\textwidth]{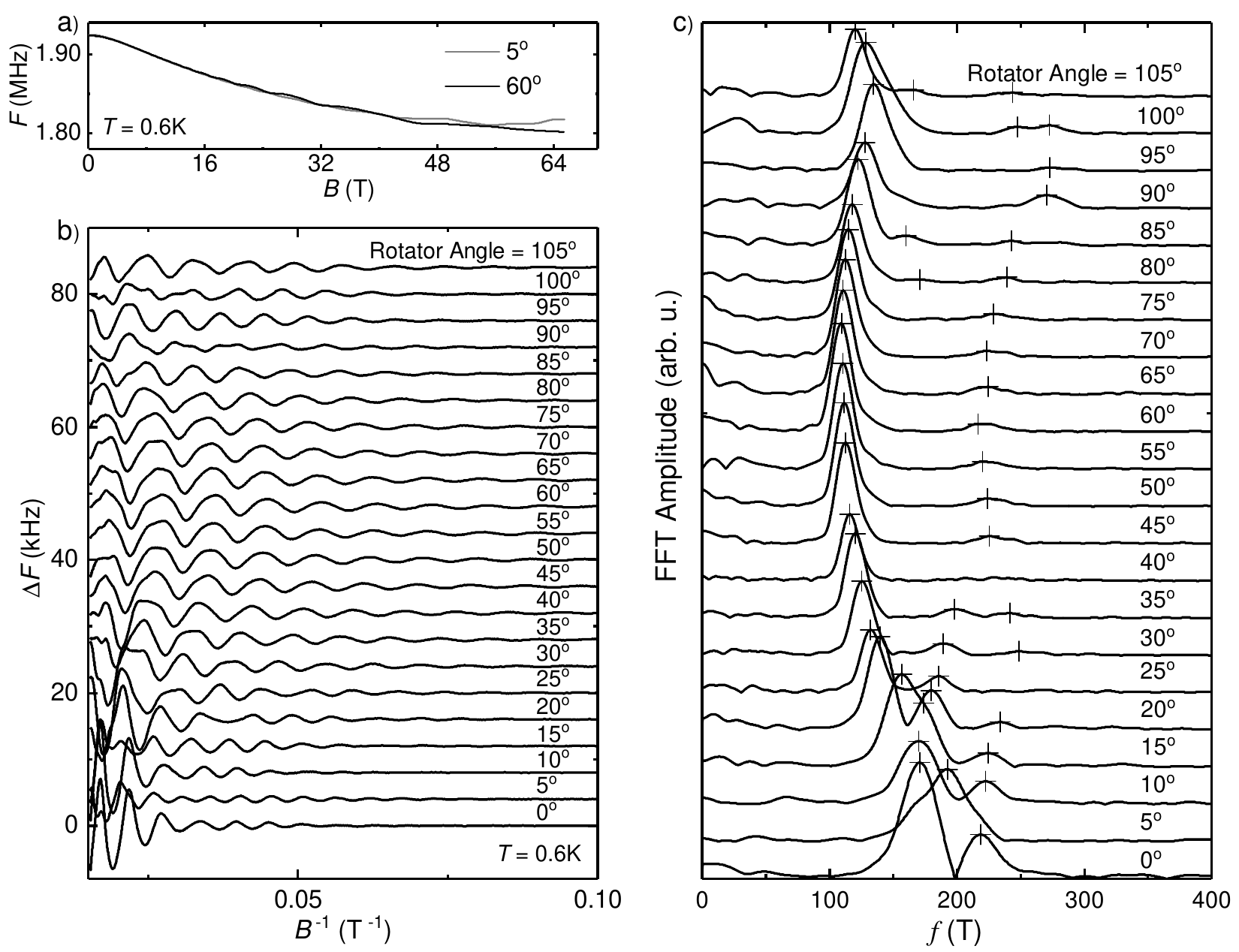}
\caption{Quantum oscillation data for a sample with $p_H(B\to 0)=$2.5$\times 10^{20}$\,cm$^{-3}$ a) Representative raw data at rotator angles of 5$^\circ$ and 60$^\circ$ expressed as the mixed-down resonant frequency, $F$, of the oscillator circuit, the change in $F$ is due to the change in conductivity of the sample mounted on an inductive coil. b) Following the subtraction of a smooth non-oscillating background, the data plotted versus inverse field shows clear periodic quantum oscillations that evolve with angle. c) Fast Fourier Transforms (FFTs) of the data in b) show the anglular dependence of the quantum oscillation frequencies, $f$, which are proportional to the area of the Fermi surface perpendicular to the applied field. Clearly identifiable frequencies are indicated with crosses.} \label{FIG:QOdata}
\end{figure*}

Representative mutual inductance data are shown in Figure \ref{FIG:QOdata}a for the sample with $p_H(B\to 0)$=2.5$\times$10$^{20}$\,cm$^{-3}$ (grey circles, Figure \ref{FIG:HallData}b) at rotator angles of 5$^\circ$ and 60$^\circ$. The sample was nominally mounted with the crystallographic [110] direction parallel to the rotator axis such that the field sweeps through the high-symmetry points of the $L$-pocket (and of the potential $\Sigma$-pocket) upon rotation. A non-oscillating background was subtracted from the data via a manual spline-fit technique (polynomial fitting yields poor results for low frequencies), the results of which are shown for all measured rotator angles in Figure \ref{FIG:QOdata}b where a clear periodicity in inverse field can be observed, indicative of quantum oscillations. The frequency spectrum is then obtained by Fast Fourier Transform (FFT) and allows the quantum oscillation frequencies to be identified as shown in Figure \ref{FIG:QOdata}c with robustly identifiable peak positions identified with crosses. The temperature and field dependences of the amplitudes of $f_{\mathrm{min}}$ and $f_{100}$ were also measured in order to determine the effective cyclotron masses, $m^*_c$, and the Dingle temperatures, $\Theta_D$ (these are presented and discussed in Appendices \ref{APP:QOmass} and \ref{APP:DingleTemp}).

\begin{figure}
\includegraphics[width=\columnwidth]{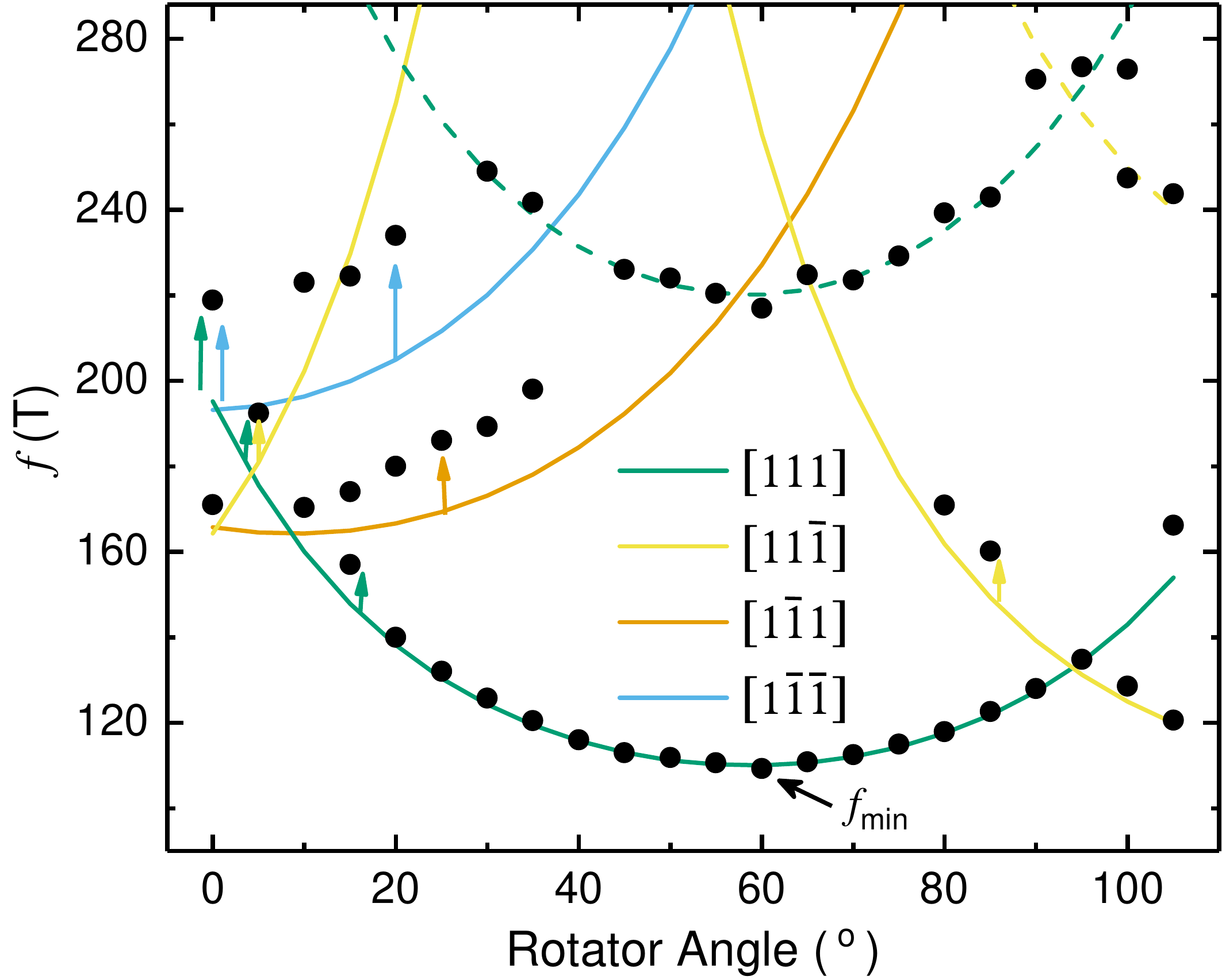}
\caption{The quantum oscillation frequencies identified in Figure \ref{FIG:QOdata} are shown versus rotator angle, with coloured lines showing the expectations of an ellipsoidal model ($K$=15). The nominal axis of rotation is [110] and 0$^\circ$ corresponds to $B \parallel$[100].  The determination of the sample misalignment required to orient the model is described in Appendix \ref{APP:ModelConstraints}. The four equivalent $L$-pockets in the model form different branches in the data due to their respective major semi-axes (indicated in the legend) being at different angles with respect to the applied magnetic field. Arrows highlight the differences between the data and the model, and the dashed lines represent higher harmonics. The uncertainty in the frequencies is approximately the size of the data points.}  \label{FIG:QOmodel}
\end{figure}

\begin{figure}
\includegraphics[width=\columnwidth]{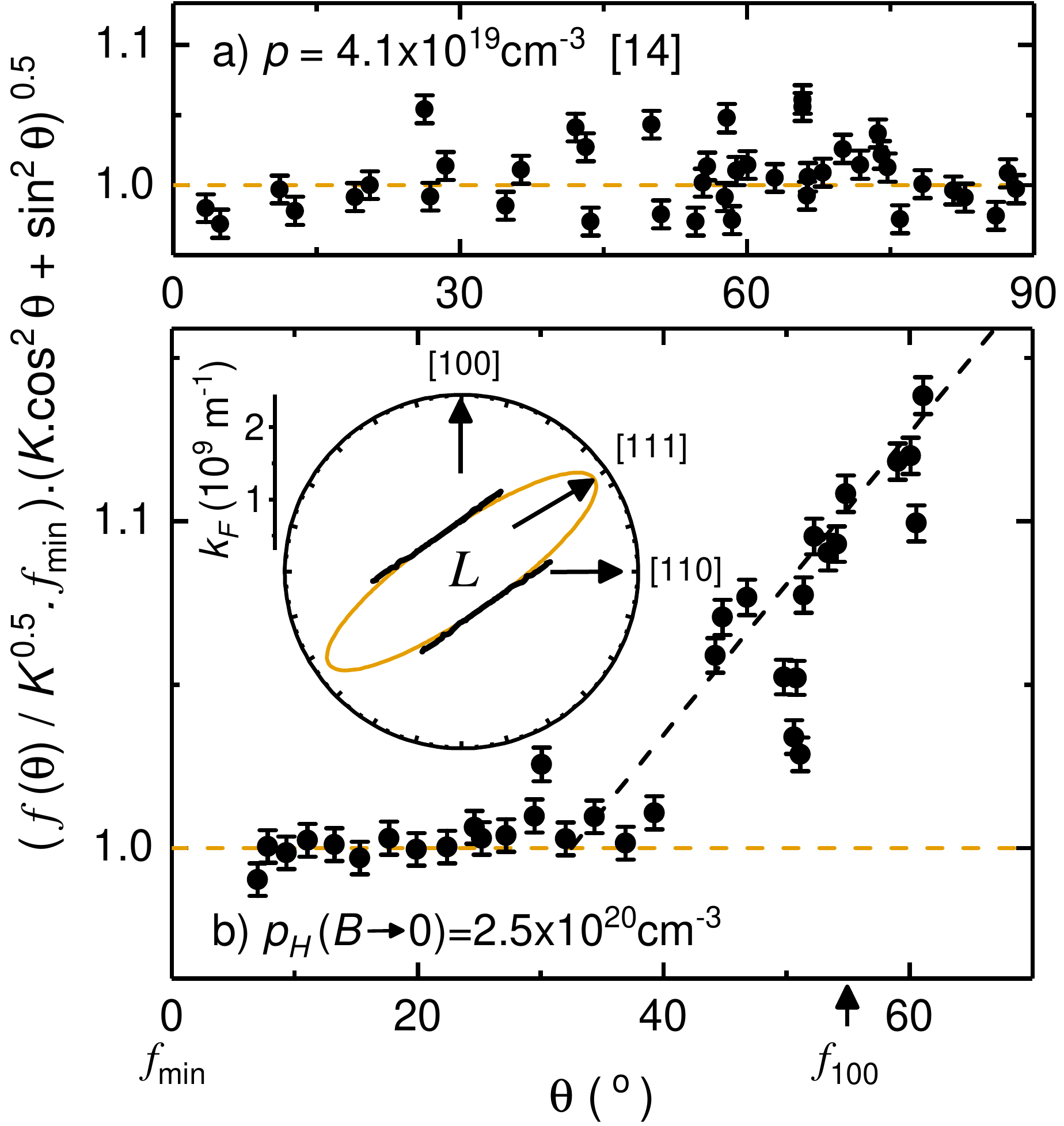}
\caption{Quantum oscillation frequencies plotted  as $(f(\theta) / K^{0.5}.f_{\mathrm{min}}).(K.\cos^2(\theta) + \sin^2(\theta))^{0.5}$ such that an ellipsoidal Fermi surface ($K=15$) yields a constant. a) shows data for a sample with $p=4.1\times 10^{19}$cm$^{-3}$ that fits the ellipsoidal model (data reanalysed from Ref [\onlinecite{Giraldo-Gallo2016}]) and b) the present data with $p_H(B \to 0) =2.5\times10^{20}$cm$^{-3}$ that shows significant deviation from the model. The ellipsoidal model is shown as a dashed orange line, with the black dashed line in b) being a guide to the eye illustrating the deviation from ellipsoidal behaviour. \emph{Inset of b):} $k_F$ derived from the data around the $L$-point of the Brillouin zone illustrating the range of the Fermi surface resolved in these measurements as well as the deviation from ellipsoidicity towards something more like a tube. The ellipsoidal model is shown in orange. } \label{FIG:QOModelPolar}
\end{figure}

Firstly, note that no evidence for an additional Fermi surface pocket was observed in either this sample or the second sample with $p_H(B\to 0)$=1.45$\times$10$^{20}$\,cm$^{-3}$ (grey triangles, Fig. \ref{FIG:HallData}b), but also that this does not rule out its existence as the $\Sigma$ pocket is anticipated to have a lower mobility and thus a lower amplitude \cite{Ravich1970}. In Figure \ref{FIG:QOmodel} we compare the extracted quantum oscillation frequencies with those expected from the Kane-type behaviour of the $L$-pocket observed at lower dopings, i.e. an ellipsoid with a fixed ellipsoidicity of $K$=15 (Appendix \ref{APP:ModelConstraints} describes how the model was constrained and sample misalignment was accounted for). The higher frequency orbits where the most pronounced changes in the shape of the $L$-pocket may be expected upon approach to the saddle point in the band structure were not resolved in the data, and so the interpretation of the deviations from the ellipsoidal model are somewhat subtle and discussed further below. The ellipsoidal model gives a reliable lower bound for the $L$-pocket Luttinger volume which is shown by the grey dot-dashed line in Figure \ref{FIG:HallData}b that falls considerably below $p_H(B\to 0)$ in the same sample, and also a lesser but possibly significant amount below the Na density of the same sample from EMPA, but outside of the high field limit we cannot compare to a good value of $p$ directly. An unexpected result is that the two samples with $p_H(B\to 0)$=1.45$\times$10$^{20}$\,cm$^{-3}$ and $p_H(B\to 0)$=2.5$\times$10$^{20}$\,cm$^{-3}$ are found to have minimum quantum oscillation frequencies, $f_{\mathrm{min}}$, that are almost identical within the experimental uncertainty, implying that this region of the Fermi surface is essentially unchanged even as $p_H(B\to 0)$ increases by a factor of around two.

The deviation of the quantum oscillation frequencies from those expected from a Kane-type, energy independent ellipsoidicity (highlighted by the arrows in Figure \ref{FIG:QOmodel}) is shown more clearly in the main panel of Figure \ref{FIG:QOModelPolar}b by plotting the data and model as $(f(\theta) / K^{0.5}.f_{\mathrm{min}}).(K.\cos^2(\theta) + \sin^2(\theta))^{0.5}$ where $\theta$ is the angle of the applied field relative to the major semi-axis of the ellipsoid to which each frequency has been attributed, and $f_{\mathrm{min}}=f(\theta=0)$ is the minimum frequency corresponding to the `belly' orbit of the $L$-pocket (located at the zone boundary and well defined by the minimum of the [111] branch). In this form the orbits from the four $L$-pockets collapse onto a single curve and the ellipsoidal model reduces to unity (orange dashed line). The validity of this analysis is confirmed in Figure \ref{FIG:QOModelPolar}a for a sample at lower doping ($p=4.1 \times 10^{19}$cm$^{-3}$) using data reanalysed from Giraldo-Gallo \emph{et al.}\cite{Giraldo-Gallo2016} where the Fermi surface is known to be ellipsoidal. Figure \ref{FIG:QOModelPolar}b shows that the present data follows the model up to $\theta \approx$30$^\circ$ but $f(\theta)$ deviates upwards at higher angles. It is important to note that simply increasing $K$ does not reconcile the model and the data for realistic values. The inset to Figure \ref{FIG:QOModelPolar}b shows the cross-section of the $L$-pocket in $k$-space deduced from this data via the Onsager relation and assuming ellipsoidal orbits ($f_{\mathrm{min}} \propto k_{\perp}^2$, $f(\theta) \propto k_{\perp}k(\theta)$), and while this assumption may not be valid for $\theta > 30^\circ$ based on the present data, this nonetheless provides a helpful illustration of the portion of the Fermi surface that has been measured and the magnitude of the apparent deviation from an ellipsoidal pocket (again shown in orange) towards something more like a tube. The excellent agreement of the harmonics in Figure \ref{FIG:QOmodel} shows that this trend is not an artifact of the background subtraction, and errors in determining the misalignment of the sample would generally both increase and decrease frequencies relative to the model, whereas in Figure \ref{FIG:QOModelPolar}b we only see a monotonic increase within the spread of the data. This data is in qualitative agreement with the shape sketched in red in Figure \ref{FIG:BSsketch}d and calculated in Figure \ref{FIG:BSsketch}d for a non-ellipsoidal $L$-pocket approaching a saddle point in the band structure, although data at larger $\theta$ where non-ellipsoidicity is more pronounced would add more weight to this conclusion.

\subsection{Specific Heat}
\begin{figure}
\includegraphics[width=\columnwidth]{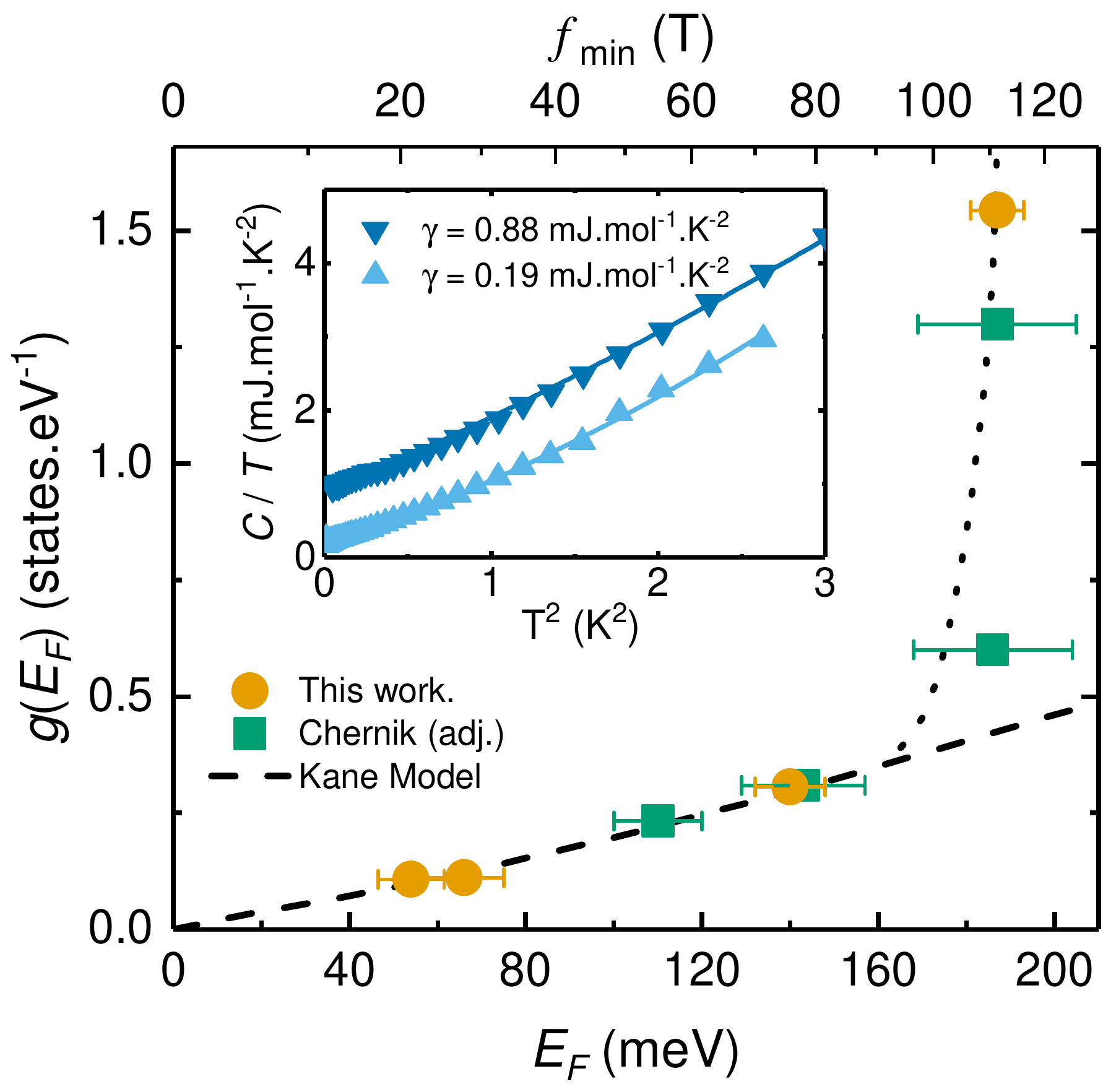}
\caption{\emph{Inset:} representative low temperature heat capacity data used to extract the Sommerfeld coefficient, $\gamma$, fitted by a second order polynomial in $C(T)/T^2$ versus $T$. \emph{Main:} the density of states at the Fermi level, $g(E_F)$, as a function of $E_F$ for the present data and data from Chernik \emph{et al}\cite{Chernik1981b}, with the relationship to $f_{\mathrm{min}}$ (from which $E_F$ is derived) also shown on a non-linear scale on the top axis. The determination of $g(E_F)$ is outlined in Appendix \ref{APP:FermiEnergy}. The data shows a sudden increase in $g(E_F)$ that onsets around 180\,meV.} \label{FIG:DoS+Cp}
\end{figure}

To learn more about the region of the Fermi surface that could not be observed by quantum oscillations we turn to specific heat measurements of the Sommerfeld coefficient, which yields the total electronic density of states at the Fermi level. Some representative data is shown in the inset to Figure \ref{FIG:DoS+Cp} illustrating the fit used to determine the Sommerfeld coefficient, which requires a second-order polynomial in $C/T$ as a function of $T^2$ in this material owing to the anomalous low-energy phonon dispersion in this material\cite{Delaire2011}. Figure \ref{FIG:DoS+Cp} shows the density of states as a function of the Fermi energy as determined from both the present and published data\cite{Giraldo-Gallo2016, Chernik1981b}, with the Fermi energy determined using $f_{\mathrm{min}}$ at high dopings. The use of $f_{\mathrm{min}}$ is important because the band structure at the Brillouin zone boundary where this orbit occcurs is expected to be unaffected  by the approach to the saddle point in the band structure and so the dispersion in this direction should give a good estimate of $E_F$ by extending the Kane model, even if the same model were to break down near the tip of the $L$-pocket. The process by which these values are combined to produce Figure \ref{FIG:DoS+Cp} is described in detail in Appendix \ref{APP:FermiEnergy}. The density of states is found to increase very rapidly as a function of energy at around 180\,meV, and much more sharply than implied by published data relying on the Hall number\cite{Chernik1981b} (see also Figure \ref{FIG:EF-DoS}) and calculations\cite{Singh2010}. This is a key result of this work and provides a new insight into the origin of the high thermopower in this material. Futhermore this provides the clearest indication that a single-band Kane type dispersion does not describe the electronic structure of Pb$_{1-x}$Na$_x$Te at the highest dopant concentrations.

\section{Discussion}
The data presented here establish two important phenomenological differences imparted by the deviation from a Kane-type dispersion that occurs around 180\,meV. Firstly, the carrier density is no longer well represented by the Hall effect at accessible magnetic fields. Secondly, the electronic density of states increases much faster as a function of energy than could be inferred from existing characterisations. The origin of this phenomenology is less immediately clear, and so here we consider the relative merits of the two Fermiological scenarios detailed in Figure \ref{FIG:BSsketch} (non-ellipsoidicity of the $L$-pocket and the presence of a pocket at $\Sigma^*$), as well as the possibility of a resonant impurity state in explaining the data.

The consideration of a resonant impurity state is motivated by how sharply the density of states seems to increase as a function of energy, which could be indicative of Fermi level pinning as has been observed previously for a number of impurities in PbTe \cite{Kaidanov1985}. Calculations that have been successful in reproducing other observed resonant impurity states in PbTe have predicted that the effect of sodium on the density of states is negligible, but also that Pb vacancies may produce a weakly interacting resonant state in this energy range\cite{Ahmad2006}. Experimentally however the absence of a signature in tunneling experiments on similarly Pb deficient crystals is notable\cite{Kaidanov1989}. Predictions of the Hall factor also suggest that the field dependence of the Hall number observed here is incompatible with a resonant impurity state as $p_H(B\to 0)<p$ in that case\cite{Prokofieva2009}, and the lack of a large increase in the Dingle temperature and residual resistivity ratio would also imply that any such state would need to be very highly localised.

It is generally accepted that there is a second valence band maximum at the $\Sigma^*$ point in PbTe, which appears 12 times in the Brillouin zone and has a much flatter dispersion than the first valence band maximum at the $L$-point. The lack of any signatures of the $\Sigma^*$ pocket in the quantum oscillation data could be understood as the mobility is expected to be lower than on the $L$-pocket, and intuitively the density of states must increase faster as a function of energy as the Fermi level is tuned below the second valence band maximum. But the two-band model of the Hall effect predicts that $p_H(B\to 0)<p$ when both bands are hole-like\footnote{The low-field {H}all coefficient modelled for two inequivalent hole-pockets is written as ${R}_{H} = \frac{1}{ec}\frac{A_{p_1}b^2_b p_{1}+A_{p_2}p_2}{(b_p p_1 +p_2)^2}$ where $p_i$ and ${A}_{p_i}$ are the hole density and {H}all factor of pocket $i$ respectively and the ratio of the mobilities of the pockets is $b_p=\frac{\mu_{p_1}}{\mu_{p_2}}$. The limiting case $\mu_{p_1}\gg \mu_{p_2}$ reduces to ${R}_{H}=\frac{A_{p_1}}{ecp_1}$, which is larger than the high-field value, ${R_H}({B} \to \infty)=\frac{1}{ec(p_1+p_2)}$ provided ${A}_{p_i}$ is sufficiently close to unity. {R}emembering that we have defined $p_{H}({B}\to0)=\frac{1}{R_H(B \to0)e}$, it is therefore the case that $p_{H}({B}\to0) < p$ for the two-band model in the absence of large anisotropies.}, in opposition to the present data\cite{Ravich1970, Hurd1972}. Elevated temperature studies of the Hall effect in PbTe confirm this intuition as $p_H(B\to 0)$ is observed to fall with increasing temperature as the $\Sigma^*$ pocket becomes populated owing to the temperature dependent band offset and thermally excited carriers\cite{Ravich1970, Allgaier1966}. This suggests that even a two-band model that accounts for the anisotropy of the $\Sigma^*$ pocket cannot reconcile the field dependence of the Hall number in this scenario.

Non-ellipsoidicity of the $L$-pocket provides the most natural mechanism  by which to explain the field-dependence of the Hall number because it can yield $p_H(B\to 0)>p$ by altering the anisotropy of the Fermi surface and the scattering rate\cite{Ong1991,Hurd1972}. Indeed the ellipsoidicity of the $L$-pocket at lower dopings already demonstrates this with a qualitatively different field dependence in samples of lower doping, as evidenced in Figure \ref{FIG:HallData}b. The most direct piece of evidence in favour of this scenario comes from the measured deviation from ellipsoidicity seen in the angle-dependence of the quantum oscillations; although the whole $L$-pocket is not resolved the effect appears to be significant in the data. The highly $\theta$ dependent loss of quantum oscillation amplitude relative to lower dopings in the absence of a significant increase in $\Theta_{D}$ (see Appendix \ref{APP:DingleTemp}) is also suggestive of increased scattering in a localised region of $k$-space around the tip of the $L$-pocket owing to increased density of states and greater phase smearing from the increased band curvature, as oppose to a second pocket or impurity states that would be expected to scatter carriers more equally around the Fermi surface. The approach to a saddle point in the band structure which would lead to non-ellipsoidicity naturally implies an increase in the rate of change of the density of states, but as with the second band maximum scenario, the feature is somewhat sharper than that expected from calculations\cite{Singh2010} and prior data\cite{Kaidanov1989}. This apparent discrepancy may be reconciled if the dispersion along the $\Lambda^*-\Sigma^*$ line defined in Figure \ref{FIG:BSsketch}d is flatter than anticipated, possibly due to a failure of the assumption of a rigid band shift upon Na doping or uncertainties in calculations of the quantitative details of the band structure of PbTe.

Regardless of its origin, the large increase in the density of states observed directly here is likely to be the one invoked when explaining the high thermopower in Pb$_{1-x}$Na$_x$Te \cite{Pei2012, Singh2010, Airapetyants1966, Prokofieva2009}, thus lending support to the argument that the high thermopower in $p$-type PbTe is intrinsic and originates from a flat feature in the valence band. The energy at which the density of states begins to increase agrees well with calculations by Singh \cite{Singh2010} that also invoke non-ellipsoidicity of the $L$-pocket as the source of the increase with the second band maximum occuring slightly lower in energy. As the same calculations have also captured the magnitude of the thermopower accurately, this seems to form a consistent picture that the phenomenology of heavily doped Pb$_{1-x}$Na$_x$Te is strongly influenced by non-ellipsoidicity of the $L$-pocket on approach to a saddle-point in the band structure.

\section{Conclusion}
In this work we have explored the electronic structure of heavily holed-doped PbTe single crystals in the ground state using the complementary probes of the Hall effect, quantum oscillations, specific heat, and EMPA, from which we identify and characterise a significant deviation from the Kane model that onsets around 180\,meV. The data show that the density of states grows sharply as a function of $E_F$ concurrently with a significant increase in the threshold at which the system enters the high-field limit as seen by the Hall effect. This leads to two important new phenomenological conclusions; firstly that the carrier density is significantly overestimated by the Hall effect at accessible fields in this regime, and secondly, that the density of states increases more rapidly with energy than previously inferred. We consider three possible sources of this phenomenology, a resonant impurity state, a second valence band maximum at the Fermi level, and non-ellipsoidicity of the $L$-pocket. By comparing the angle-dependence of the quantum oscillation measurements to an ellipsoidal model and considering the field dependence of the Hall number we propose that non-ellipsoidicity of the $L$-pocket is the most likely origin of this phenomenology, and this scenario is also most consistent with density functional theory. In order to reconcile this scenario with the rapid increase in the density of states, the band dispersion along the $\Sigma^*-\Lambda^*$ line (defined in Figure \ref{FIG:BSsketch}d) must be flatter than anticipated from calculations. These results contribute to a consistent picture of the origin of the high thermopower of Pb$_{1-x}$Na$_x$Te, as well as calling for a reassessment of the real carrier densities in heavily doped PbTe.

\section{Acknowledgements}
The authors would like to thank Johanna Palmstrom and Scott Hannahs for experimental support at the NHMFL in Tallahassee and Filip Ronning for assistance with additional sample characterisation at Los Alamos. We thank Dale Burns for assistance with EMPA measurements at the Stanford Mineral and Microchemical Analysis Facility, and Heike Pfau for pursuing additional characterisation. Calculated band structures were kindly provided by Nicola Spaldin and Boris Sangiorgio. P.W. and I.R.F. were supported by AFOSR Grant No. FA9550-09-1-0583. PW acknowledges partial support from the Gordon and Betty Moore Foundations EPiQS Initiative through grant GBMF4414. The National High Magnetic Field Laboratory is supported by the National Science Foundation through NSF/DMR-1644779 and DMR-1157490, and the State of Florida. The Pulsed Field Facility at Los Alamos National Labratory was additionally supported by the the U.S. Department of Energy.

\section{Appendices}
\begin{appendices}

\subsection{Temperature depence of quantum oscillations and estimates of $m^*_c$}
\label{APP:QOmass}
\begin{figure}
\includegraphics[width=\columnwidth]{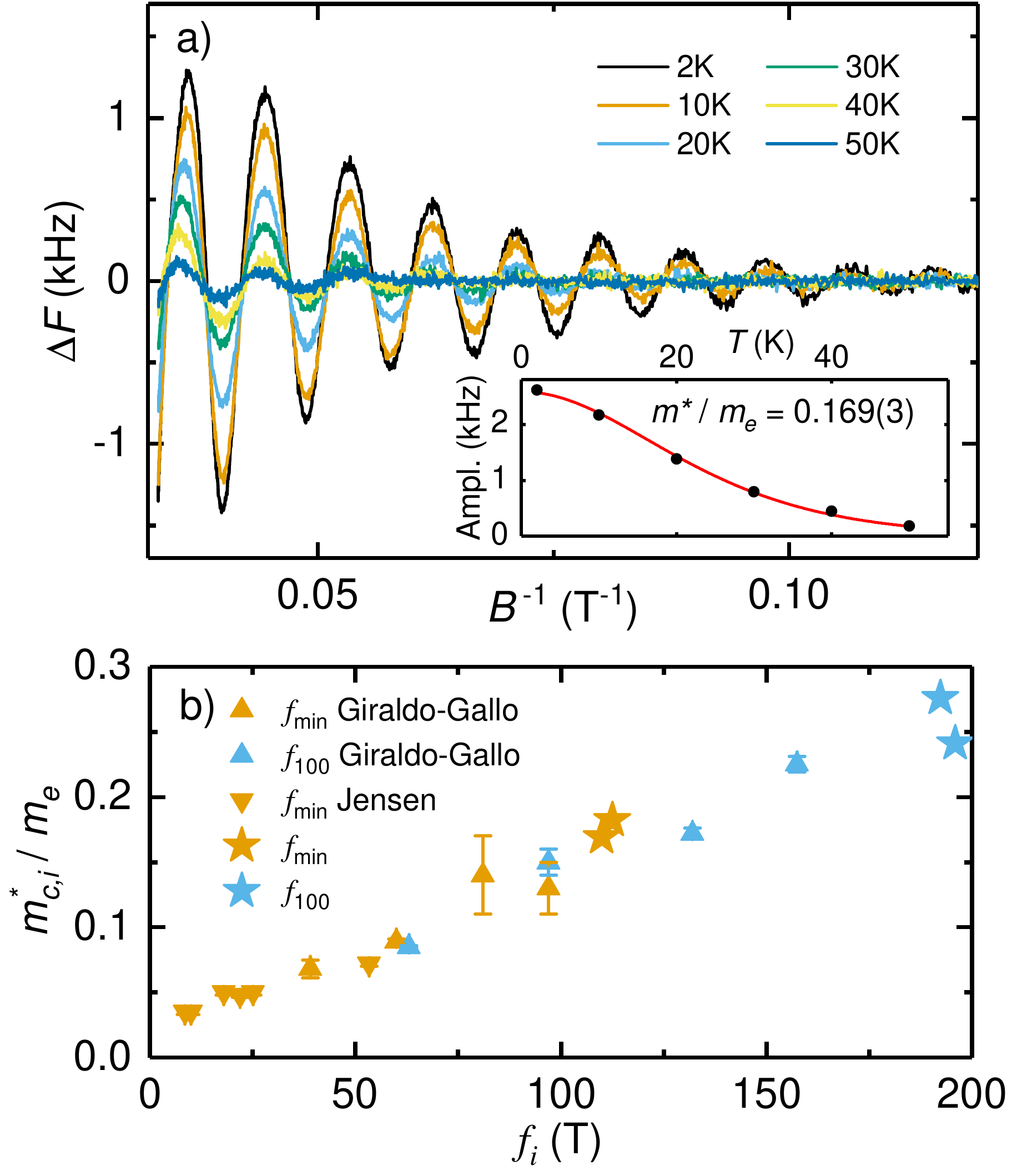}
\caption{a) Representative effective mass data showing the reduction of the quantum oscillation ampitude with temperature in the sample with $p_H(B\to 0)$=2.5$\times$10$^{20}$\,cm$^{-3}$, shown here for the $f_{\mathrm{min}}$ ($B \parallel [111]$) orbit. \emph{Inset:} shows a fit to the temperature dependent term of the Lifshitz-Kosevitch formula that yields the cyclotron effective mass, $m^*_c$=0.169(3)$\,m_e$. b) shows the relationship between $m^*_{c,\mathrm{min}}$ and $f_{\mathrm{min}}$ and the relationship between $m^*_{c,100}$ and $f_{100}$ including data from Giraldo-Gallo \emph{et al.}\cite{Giraldo-Gallo2016} and Jensen \emph{et al.}\cite{Jensen1978}.} \label{FIG:QOmass}
\end{figure}

The temperature dependence of the amplitude of the $f_{\mathrm{min}}$ orbit ($B \parallel [111]$) is shown in Figure \ref{FIG:QOmass}a for the sample with $p_H(B\to 0)$=2.5$\times$10$^{20}$\,cm$^{-3}$. By fitting the amplitude of the oscillation as a function of temperature to the temperature dependent term of the Lifshitz-Kosevitch formula, as shown in the inset to Figure \ref{FIG:QOmass}a, we obtain the cyclotron effective mass, $m_c^*$=0.169(3)$\,m_e$ for the $f_{\mathrm{min}}$ orbit. Similar data was taken for $f_{100}$ and also in sample with $p_H(B\to 0)$=1.45$\times$10$^{20}$\,cm$^{-3}$, and these results are compared to data from Giraldo-Gallo \emph{et al.}\cite{Giraldo-Gallo2016} in Figure \ref{FIG:QOmass}b as function of the orbit frequency. This yields a linear trend as expected from the Kane model where $m^*_{c,i}\propto E_F \propto k_{F,i}^2 \propto f_i$, with the $f_{\mathrm{min}}$ and $f_{100}$ data collapsing onto the same curve because the cross sectional area of the Fermi surface and the cyclotron mass share the same angular dependence\cite{Ravich1970}. Given the apparent deviation from ellipsoidicity highlighted in Figure \ref{FIG:QOModelPolar}b one may expect an increase in the effective mass in the present samples for the (100) orbit, however $m^*_c$ represents an average of $m^*$ around the completed orbit, and so $m^*_c$ is expected to see a smaller enhancement than is present in the portion of the orbit with the most band curvature. Furthermore, any increase would in part be cancelled out by the concurrent increase in the frequency when presented as in this plot, and so it is not clear that $m^*_c$ constitutes a particularly sensitive probe of non-ellipsoidicity in this context. The close adherence of the $f_{\mathrm{min}}$ data to the Kane model justifies the use of $f_{\mathrm{min}}$ in determining the Fermi energy as discussed below in Appendix \ref{APP:FermiEnergy}.

\subsection{Field dependence of quantum oscillations and the Dingle temperature}
\begin{figure}
\includegraphics[width=\columnwidth]{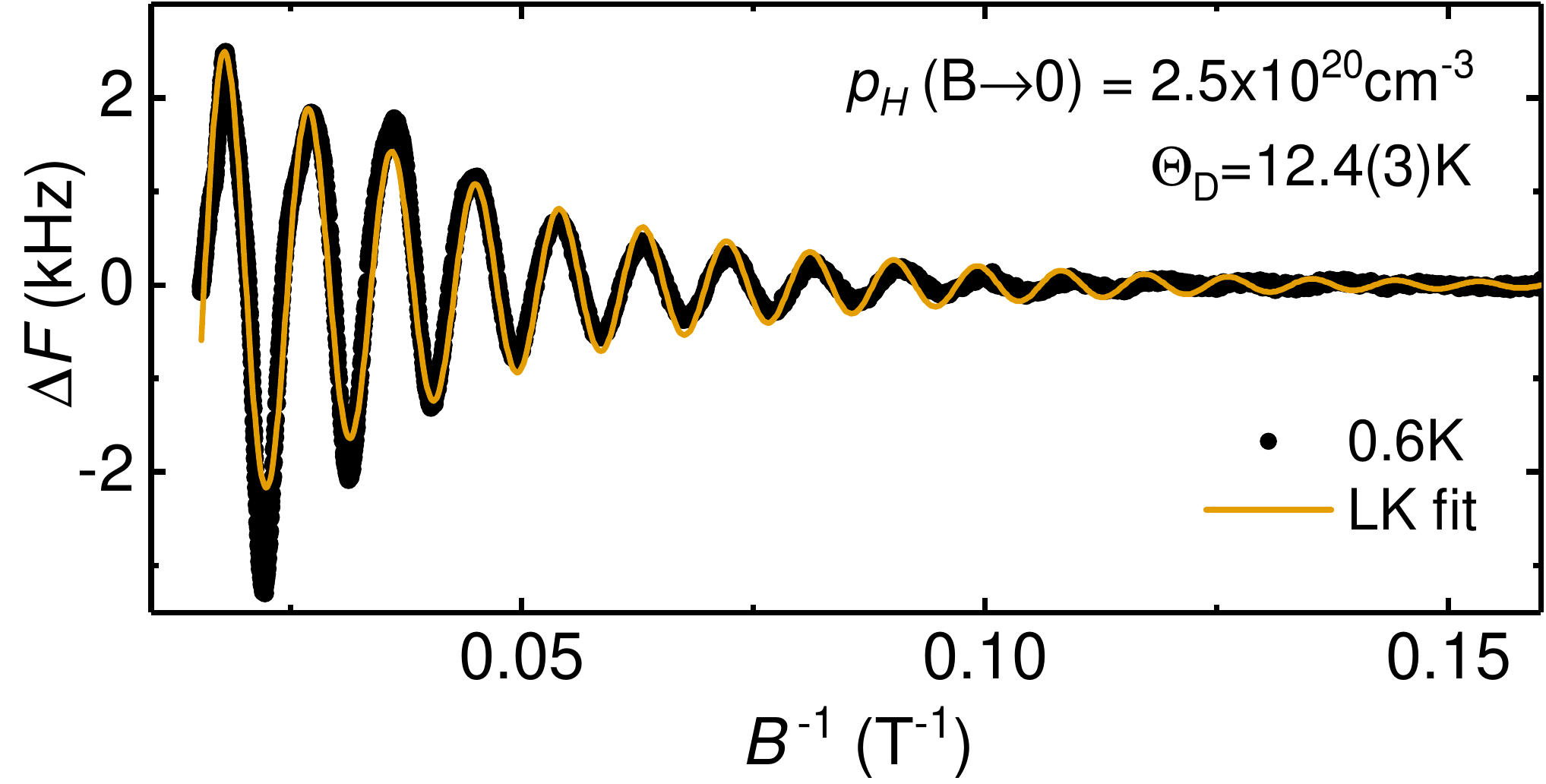}
\caption{A Lifshitz-Kosevitch fit to data (orange line) at $f_{\mathrm{min}}$ ($B \parallel [111]$) at 0.6\,K in the $p_H(B\to0)=2.5\times10^{20}$\,cm$^{-3}$ sample, yielding a Dingle temperature of $\Theta_{D}$=12.4(3)\,K. A similar fit to the $p_H(B\to0)=1.45\times10^{20}$\,cm$^{-3}$ sample gives $\Theta_{D}$=12.3(3)\,K.} \label{FIG:QOdingle}
\end{figure}

The magnetic field dependence of a quantum oscillation amplitude is related by the Dingle temperature, $\Theta_D$, to the mean free path. It is curious that Giraldo-Gallo \emph{et al.}\cite{Giraldo-Gallo2016} found that $\Theta_D$ is independent of doping in Pb$_{1-x}$Na$_x$Te, when the mean free path might be expected to fall as more dopants are introduced, which may reflect the very strong screening effect in PbTe. Figure \ref{FIG:QOdingle} shows a fit of the Lifshitz-Kosevitch formula which determines $\Theta_{D}$ to be 12.4(3)\,K for the $f_{\mathrm{min}}$ orbit, slightly larger but comparable to that at lower dopings. This comparison is complicated slightly as the samples used here are grown by a melt-growth method that is likely to be less structurally perfect than the vapor-transport method used at lower dopings by Giraldo-Gallo \emph{et al.} \cite{Kokhlov2003}. It is notable however that a resonant impurity state might be expected to have a larger effect on $\Theta_D$ than this, as observed in the strong suppression of the quantum oscillation amplitude when doping with Tl beyond 0.3\% \cite{Giraldo-Gallo2017}.
\label{APP:DingleTemp}

\subsection{Constraining the Ellipsoidal Model}
\label{APP:ModelConstraints}
The model shown in in Figure \ref{FIG:QOmodel} was constrained in the following ways; the ellipsoidicity was held as $K=15$ for comparison to lower carrier concentrations, noting that the fit is largely insensitive to realistic changes of this value for the observed frequencies; the minimum quantum oscillation frequency, $f_{\mathrm{min}}$, corresponding to the cross-sectional area of the $L$-pocket at the zone boundary, was taken as accurate because this value changes very little with misalignment; the misalignment and the uncertainty in the misalignment were estimated by optical measurements of the experimental setup; the quality of fit of the model was assessed within the constraints of the uncertainty in the misalignment by a least-squares type analysis, the conclusion of which was that there was no realistic combination of parameters by which the ellipsoidal model could fit all of the data simultaneously. The displayed model is that determined by the optically estimated misalignments, and, as discussed in the main text, is consistent with how the $L$-pocket may be expected to deviate from ellipsoidicity. Two empirical statements support the validity of the final parameters of the model and the resulting comparison to the analysed data, firstly the harmonics seem to fit very well at low $\mathrm{\theta}$ implying that any deviation is as a function of $\mathrm{\theta}$, not a function of the measured $f(T)$ as may be expected from a systematic error. Also, in order to improve the fit to either of the [1$\overbar{1}$1] and [1$\overbar{1}\overbar{1}$] branches, the fit to the other must be compromised. Put differently, the non-misaligned [1$\overbar{1}$1] and [1$\overbar{1}\overbar{1}$] branches would be approximately an average of the two misaligned branches, and it can be seen that the average of the two measured branches is significantly higher than those of the model. The effect of a badly estimated misalignment would produce values both above and below unity in Figure \ref{FIG:QOModelPolar}, and indeed this may contribute to some of the additional scatter in Figure \ref{FIG:QOModelPolar}a, but cannot explain the behaviour observed in Figure \ref{FIG:QOModelPolar}b.

\subsection{Determination of $g(E_F)$ }
\label{APP:FermiEnergy}

\begin{figure}
\includegraphics[width=\columnwidth]{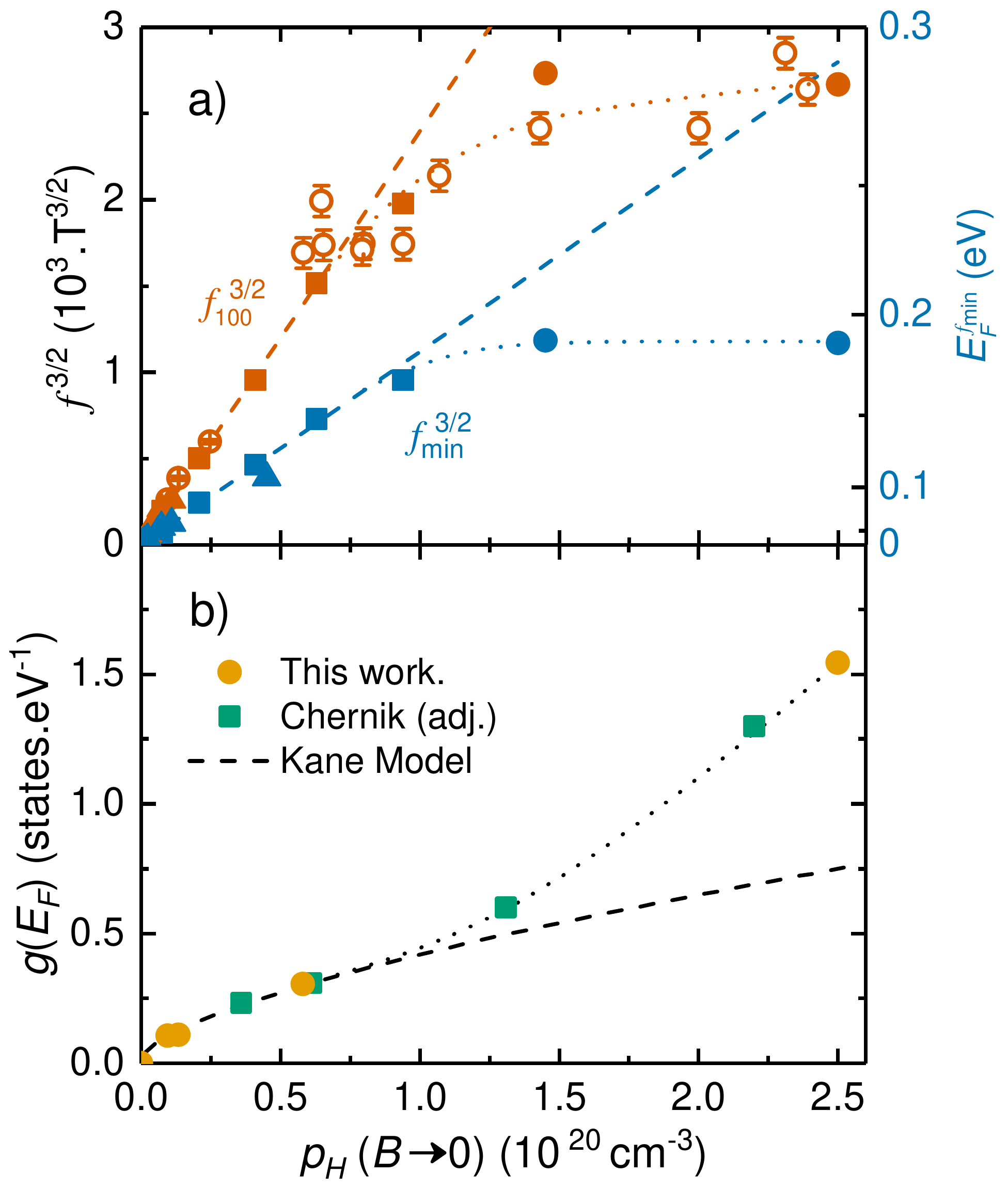}
\caption{Data showing the process by which $g(E_F)$ and $E_F$ can be correlated in Pb$_{1-x}$Na$_x$Te via $p_H(B \to 0)$.  a) shows the evolution of $f_{\mathrm{min}}^{3/2}$ (blue) and $f_{100}^{3/2}$ (red) as a function of $p_H(B \to 0)$. Pulsed field measurements up to 65\,T and DC field measurements up to 14\,T from this work are shown as closed and open circles respectively, squares are Pb$_{1-x}$Na$_x$Te samples from Giraldo-Gallo \emph{et al.} \cite{Giraldo-Gallo2016} and triangles are Pb-vacancy doped samples from Jensen \emph{et al}\cite{Jensen1978}. The Kane model is shown as dashed lines, and a clear deviation is seen at $p_H(B \to 0) \approx 9\times 10^{19}$\,cm$^{-3}$ where both frequencies saturate. $E_F$ can be calculated from the Kane model using $f_{\mathrm{min}}$ with the conversion given by the second $y$-axis showing $E^{f_{\mathrm{min}}}_F$ (note that this scale is non-linear). b) $g(E_F)$ as determined by heat capacity data in this work (orange circles) and by Chernik \emph{et al.}\cite{Chernik1981b} (green squares) plotted as a function of $p_H(B \to 0)$. Data from Chernik \emph{et al.} has been adjusted to account for inequivalency in the methodologies as described in the text.} \label{FIG:EF-DoS}
\end{figure}

Given that the Hall number cannot be taken as a good estimate of the carrier density in the heavily doped regime, we turn to the quantum oscillation frequencies to determine $E_F$. $f_{\mathrm{min}}$ is not expected to deviate markedly from the Kane model until much larger values of $E_F$, and Figures \ref{FIG:QOModelPolar}b and \ref{FIG:QOmass}b show that this is true for all orbits up to $\mathrm{\theta} =30^{\circ}$ even in our most heavily doped sample. Figure \ref{FIG:EF-DoS}a demonstrates the expected $p_H(B\to0) \propto f^{3/2}$ (dashed line) for the lower doping regime where the Kane model holds and  $p_H(B\to0) \approx p$, as well as the strong deviation from this trend above $p_H(B\to0)\approx 9\times10^{19}$\,cm$^{-3}$. The right-hand scale shows how the $f_{\mathrm{min}}$ data points (blue) then translate to $E_F$ (non-linear scale) according to a Kane model that should remain valid for the $f_{\mathrm{min}}$ orbit. Figure \ref{FIG:EF-DoS}b also includes $f_{100}$ data including melt-grown samples that were only measured to lower fields (14T) in a single orientation. This serves to highlight the general phenomenology with a higher data density as the $f_{\mathrm{min}}$ data is quite broadly spaced (the samples naturally cleave in the (100) plane so this is the alignment used for basic characterisation, whereas a rotator is required to access $f_{\mathrm{min}}$).

The total electronic density of states, $g(E_F)$, is determined from the Sommerfeld coefficient of the specific heat, and in Figure \ref{FIG:EF-DoS}b we again compare to $p_H(B\to 0)$ which is available for all of our samples and also published data by Chernik \emph{et al.} The difficulty with measuring the Sommerfeld coefficient in PbTe is both the small density of states and the low-energy non-linearity of the acoustic phonon branch \cite{Delaire2011} that make traditional extrpolations of the $C_p=\gamma T + \beta_1 T^3$ type unsuitable. The methodology used by Chernik \emph{et al.} differs slightly from that used here but the two are complementary. Here we used lower temperatures to limit the extrapolation error and allowed a $\beta_2 T^5$ contribution to $C_{p}$ in the fit, as shown in the inset of Figure \ref{FIG:DoS+Cp}. Chernik \emph{et al.} instead used a nominally undoped PbTe as a background and subtracted this data from that of their doped samples on the assumption that only the $\gamma T$ term changes. However PbTe doesn't form without some carriers due to vacancies, stated by Chernik \emph{et al.} to be of order 10$^{18}$\,cm$^{-3}$ in their reference sample, meaning that $\gamma$ is underestimated in their data. From the Kane model we can estimate the missing $\gamma$ due to their imperfect background subtraction to be approximately 0.04\,mJ.mol$^{-1}$.K$^{-2}$, which, as expected, brings both datasets into perfect agreement. With both $E_F$ and $g(E_F)$ now on a common axis, $p_H(B \to 0)$, they can be correlated to produce Figure \ref{FIG:DoS+Cp}. Whilst there is some interpolation involved in this process, the qualitative behaviour is absolutely clear: as $f_{\mathrm{min}}$, and hence $E_F$, stops increasing, $g(E_F)$ rises significantly.

\begin{figure}
\includegraphics[width=\columnwidth]{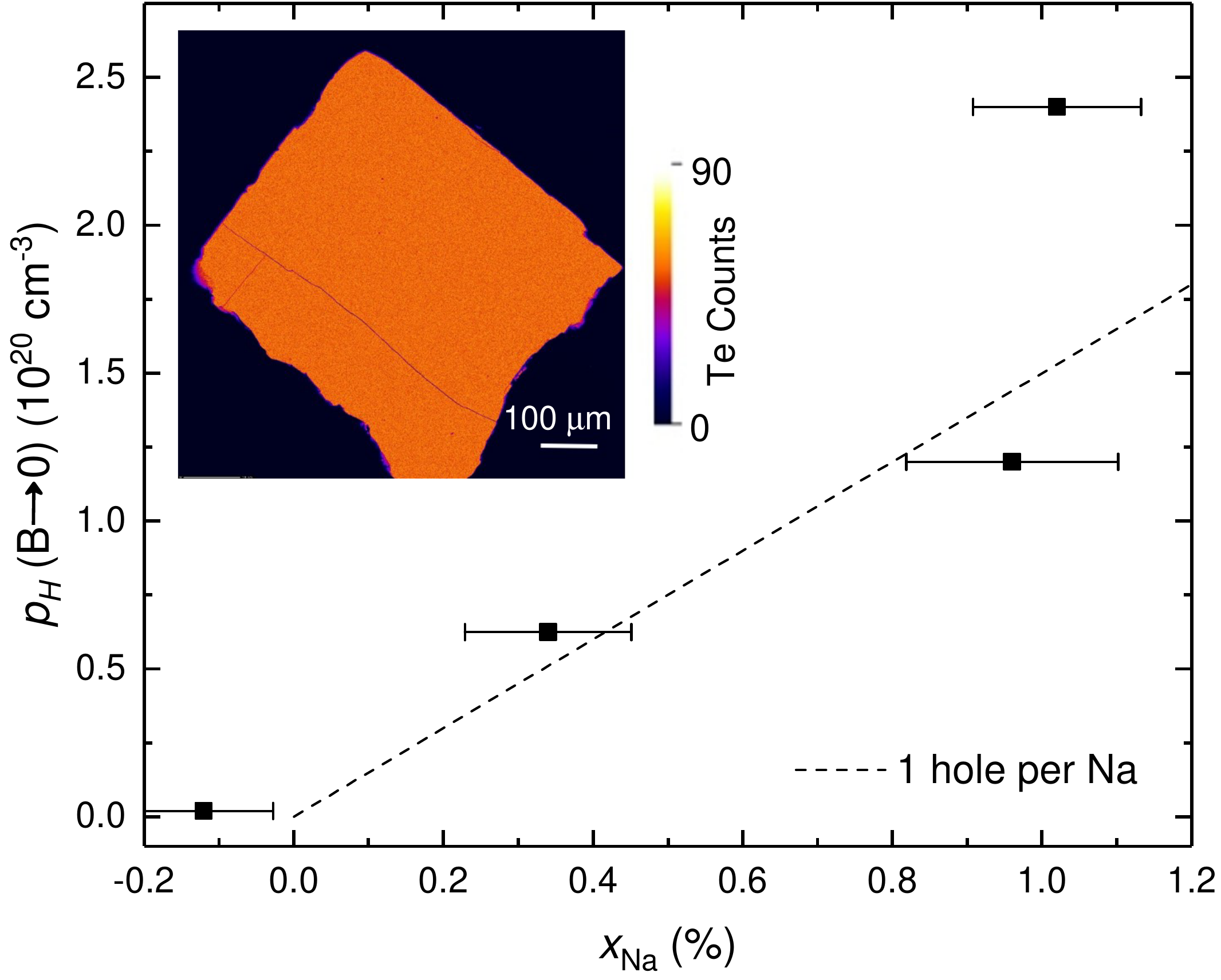}
\caption{\emph{Main:} $p_H(B \to 0)$ versus mean Na content as determined by EMPA. The dashed line shows the expectation if there is one carrier per dopant. The two highest dopings were measured on the exact same samples as the present quantum oscillation and Hall measurements, with the lower two taken from the same batches as samples measured by Giraldo-Gallo \emph{et al.} \emph{Inset:} a typical composition map, here showing Te counts in the $p_H(B \to 0)=$2.5$\times 10^{20}$\,cm$^{-3}$ sample, demonstrating that there are no impurity phases in these single crystals. Note that the lines are cracks that appeared upon polishing the sample for this measurement.} \label{FIG:EMPA}
\end{figure}

\subsection{Electron Microprobe Analysis}
\label{APP:EMPA}

As a cross-check for our conclusion that  $p_H(B \to 0) \neq p$ we performed a direct chemical analysis by EMPA. Na is known to contribute a single hole to PbTe and as a member of group 1 it is not disposed to alternative valences, and so we can assume that this behaviour does not change (unlike say, Tl which has been shown to change valence in PbTe as a function of $E_F$)\cite{Nemov1998,Giraldo-Gallo2017,Walmsley2018}. Hence the real carrier density should be close to the real dopant density provided that $n_{Na}\gg n_{vacancies}$ as expected here. Figure \ref{FIG:EMPA} confirms that  $p_H(B \to 0)$ and $n_{Na}$ (plotted as $x$) agree well at lower dopings but then deviate for the highest dopings, with the solubility limit (highest doping) seen to be approximately 1\%, corresponding to an implied carrier density of around 1.5(2)$\times$10$^{20}$\,cm$^{-3}$. This number and uncertainty are contextualised against the Hall number and lower-bound of the $L$-pocket Luttinger volume in Figure \ref{FIG:HallData}b, shown as the grey bar. The inset to Figure \ref{FIG:EMPA} shows a typical composition map, chosen here for Te to demonstrate the absence of NaTe$_2$, illustrating that there are no resolvable impurity phases in these single crystals. The Pb and Na maps look essentially identical to the Te map shown. Note that the apparently negative value of $x_{Na}$ for one data point in Figure \ref{FIG:EMPA} is an artifact of the background subtraction (the sample was undoped, i.e. $x_{Na}=0$).

\end{appendices}


\begin{thebibliography}{39}%
\makeatletter
\providecommand \@ifxundefined [1]{%
 \@ifx{#1\undefined}
}%
\providecommand \@ifnum [1]{%
 \ifnum #1\expandafter \@firstoftwo
 \else \expandafter \@secondoftwo
 \fi
}%
\providecommand \@ifx [1]{%
 \ifx #1\expandafter \@firstoftwo
 \else \expandafter \@secondoftwo
 \fi
}%
\providecommand \natexlab [1]{#1}%
\providecommand \enquote  [1]{``#1''}%
\providecommand \bibnamefont  [1]{#1}%
\providecommand \bibfnamefont [1]{#1}%
\providecommand \citenamefont [1]{#1}%
\providecommand \href@noop [0]{\@secondoftwo}%
\providecommand \href [0]{\begingroup \@sanitize@url \@href}%
\providecommand \@href[1]{\@@startlink{#1}\@@href}%
\providecommand \@@href[1]{\endgroup#1\@@endlink}%
\providecommand \@sanitize@url [0]{\catcode `\\12\catcode `\$12\catcode
  `\&12\catcode `\#12\catcode `\^12\catcode `\_12\catcode `\%12\relax}%
\providecommand \@@startlink[1]{}%
\providecommand \@@endlink[0]{}%
\providecommand \url  [0]{\begingroup\@sanitize@url \@url }%
\providecommand \@url [1]{\endgroup\@href {#1}{\urlprefix }}%
\providecommand \urlprefix  [0]{URL }%
\providecommand \Eprint [0]{\href }%
\providecommand \doibase [0]{http://dx.doi.org/}%
\providecommand \selectlanguage [0]{\@gobble}%
\providecommand \bibinfo  [0]{\@secondoftwo}%
\providecommand \bibfield  [0]{\@secondoftwo}%
\providecommand \translation [1]{[#1]}%
\providecommand \BibitemOpen [0]{}%
\providecommand \bibitemStop [0]{}%
\providecommand \bibitemNoStop [0]{.\EOS\space}%
\providecommand \EOS [0]{\spacefactor3000\relax}%
\providecommand \BibitemShut  [1]{\csname bibitem#1\endcsname}%
\let\auto@bib@innerbib\@empty
\bibitem [{\citenamefont {Pei}\ \emph {et~al.}(2012)\citenamefont {Pei},
  \citenamefont {Wang},\ and\ \citenamefont {Snyder}}]{Pei2012}%
  \BibitemOpen
  \bibfield  {author} {\bibinfo {author} {\bibfnamefont {Y.}~\bibnamefont
  {Pei}}, \bibinfo {author} {\bibfnamefont {H.}~\bibnamefont {Wang}}, \ and\
  \bibinfo {author} {\bibfnamefont {G.~J.}\ \bibnamefont {Snyder}},\ }\href
  {\doibase 10.1002/adma.201202919} {\bibfield  {journal} {\bibinfo  {journal}
  {Advanced Materials}\ }\textbf {\bibinfo {volume} {24}},\ \bibinfo {pages}
  {6125} (\bibinfo {year} {2012})}\BibitemShut {NoStop}%
\bibitem [{\citenamefont {Biswas}\ \emph {et~al.}(2012)\citenamefont {Biswas},
  \citenamefont {He}, \citenamefont {Blum}, \citenamefont {Wu}, \citenamefont
  {Hogan}, \citenamefont {Seidman}, \citenamefont {Dravid},\ and\ \citenamefont
  {Kanatzidis}}]{Biswas2012}%
  \BibitemOpen
  \bibfield  {author} {\bibinfo {author} {\bibfnamefont {K.}~\bibnamefont
  {Biswas}}, \bibinfo {author} {\bibfnamefont {J.}~\bibnamefont {He}}, \bibinfo
  {author} {\bibfnamefont {I.~D.}\ \bibnamefont {Blum}}, \bibinfo {author}
  {\bibfnamefont {C.~I.}\ \bibnamefont {Wu}}, \bibinfo {author} {\bibfnamefont
  {T.~P.}\ \bibnamefont {Hogan}}, \bibinfo {author} {\bibfnamefont {D.~N.}\
  \bibnamefont {Seidman}}, \bibinfo {author} {\bibfnamefont {V.~P.}\
  \bibnamefont {Dravid}}, \ and\ \bibinfo {author} {\bibfnamefont {M.~G.}\
  \bibnamefont {Kanatzidis}},\ }\href {\doibase 10.1038/nature11439} {\bibfield
   {journal} {\bibinfo  {journal} {Nature}\ }\textbf {\bibinfo {volume}
  {489}},\ \bibinfo {pages} {414} (\bibinfo {year} {2012})}\BibitemShut
  {NoStop}%
\bibitem [{\citenamefont {Wu}\ \emph {et~al.}(2014)\citenamefont {Wu},
  \citenamefont {Zhao}, \citenamefont {Zheng}, \citenamefont {Wu},
  \citenamefont {Pei}, \citenamefont {Tong}, \citenamefont {Kanatzidis},\ and\
  \citenamefont {He}}]{Wu2014}%
  \BibitemOpen
  \bibfield  {author} {\bibinfo {author} {\bibfnamefont {H.~J.}\ \bibnamefont
  {Wu}}, \bibinfo {author} {\bibfnamefont {L.~D.}\ \bibnamefont {Zhao}},
  \bibinfo {author} {\bibfnamefont {F.~S.}\ \bibnamefont {Zheng}}, \bibinfo
  {author} {\bibfnamefont {D.}~\bibnamefont {Wu}}, \bibinfo {author}
  {\bibfnamefont {Y.~L.}\ \bibnamefont {Pei}}, \bibinfo {author} {\bibfnamefont
  {X.}~\bibnamefont {Tong}}, \bibinfo {author} {\bibfnamefont {M.~G.}\
  \bibnamefont {Kanatzidis}}, \ and\ \bibinfo {author} {\bibfnamefont {J.~Q.}\
  \bibnamefont {He}},\ }\href {\doibase 10.1038/ncomms5515} {\bibfield
  {journal} {\bibinfo  {journal} {Nature Communications}\ }\textbf {\bibinfo
  {volume} {5}},\ \bibinfo {pages} {4515} (\bibinfo {year} {2014})}\BibitemShut
  {NoStop}%
\bibitem [{\citenamefont {Zhao}\ \emph {et~al.}(2013)\citenamefont {Zhao},
  \citenamefont {Wu}, \citenamefont {Hao}, \citenamefont {Wu}, \citenamefont
  {Zhou}, \citenamefont {Biswas}, \citenamefont {He}, \citenamefont {Hogan},
  \citenamefont {Uher}, \citenamefont {Wolverton}, \citenamefont {Dravid},\
  and\ \citenamefont {Kanatzidis}}]{Zhao2013}%
  \BibitemOpen
  \bibfield  {author} {\bibinfo {author} {\bibfnamefont {L.~D.}\ \bibnamefont
  {Zhao}}, \bibinfo {author} {\bibfnamefont {H.~J.}\ \bibnamefont {Wu}},
  \bibinfo {author} {\bibfnamefont {S.~Q.}\ \bibnamefont {Hao}}, \bibinfo
  {author} {\bibfnamefont {C.~I.}\ \bibnamefont {Wu}}, \bibinfo {author}
  {\bibfnamefont {X.~Y.}\ \bibnamefont {Zhou}}, \bibinfo {author}
  {\bibfnamefont {K.}~\bibnamefont {Biswas}}, \bibinfo {author} {\bibfnamefont
  {J.~Q.}\ \bibnamefont {He}}, \bibinfo {author} {\bibfnamefont {T.~P.}\
  \bibnamefont {Hogan}}, \bibinfo {author} {\bibfnamefont {C.}~\bibnamefont
  {Uher}}, \bibinfo {author} {\bibfnamefont {C.}~\bibnamefont {Wolverton}},
  \bibinfo {author} {\bibfnamefont {V.~P.}\ \bibnamefont {Dravid}}, \ and\
  \bibinfo {author} {\bibfnamefont {M.~G.}\ \bibnamefont {Kanatzidis}},\ }\href
  {\doibase 10.1039/c3ee42187b} {\bibfield  {journal} {\bibinfo  {journal}
  {Energy and Environmental Science}\ }\textbf {\bibinfo {volume} {6}},\
  \bibinfo {pages} {3346} (\bibinfo {year} {2013})}\BibitemShut {NoStop}%
\bibitem [{\citenamefont {Heremans}\ \emph {et~al.}(2008)\citenamefont
  {Heremans}, \citenamefont {Jovovic}, \citenamefont {Toberer}, \citenamefont
  {Saramat}, \citenamefont {Kurosaki}, \citenamefont {Charoenphakdee},
  \citenamefont {Yamanaka},\ and\ \citenamefont {Snyder}}]{Heremans2008}%
  \BibitemOpen
  \bibfield  {author} {\bibinfo {author} {\bibfnamefont {J.~P.}\ \bibnamefont
  {Heremans}}, \bibinfo {author} {\bibfnamefont {V.}~\bibnamefont {Jovovic}},
  \bibinfo {author} {\bibfnamefont {E.~S.}\ \bibnamefont {Toberer}}, \bibinfo
  {author} {\bibfnamefont {A.}~\bibnamefont {Saramat}}, \bibinfo {author}
  {\bibfnamefont {K.}~\bibnamefont {Kurosaki}}, \bibinfo {author}
  {\bibfnamefont {A.}~\bibnamefont {Charoenphakdee}}, \bibinfo {author}
  {\bibfnamefont {S.}~\bibnamefont {Yamanaka}}, \ and\ \bibinfo {author}
  {\bibfnamefont {G.~J.}\ \bibnamefont {Snyder}},\ }\href {\doibase
  10.1126/science.1159725} {\bibfield  {journal} {\bibinfo  {journal}
  {Science}\ }\textbf {\bibinfo {volume} {321}},\ \bibinfo {pages} {554}
  (\bibinfo {year} {2008})}\BibitemShut {NoStop}%
\bibitem [{\citenamefont {Delaire}\ \emph {et~al.}(2011)\citenamefont
  {Delaire}, \citenamefont {Ma}, \citenamefont {Marty}, \citenamefont {May},
  \citenamefont {McGuire}, \citenamefont {Du}, \citenamefont {Singh},
  \citenamefont {Podlesnyak}, \citenamefont {Ehlers}, \citenamefont {Lumsden},\
  and\ \citenamefont {Sales}}]{Delaire2011}%
  \BibitemOpen
  \bibfield  {author} {\bibinfo {author} {\bibfnamefont {O.}~\bibnamefont
  {Delaire}}, \bibinfo {author} {\bibfnamefont {J.}~\bibnamefont {Ma}},
  \bibinfo {author} {\bibfnamefont {K.}~\bibnamefont {Marty}}, \bibinfo
  {author} {\bibfnamefont {A.~F.}\ \bibnamefont {May}}, \bibinfo {author}
  {\bibfnamefont {M.~A.}\ \bibnamefont {McGuire}}, \bibinfo {author}
  {\bibfnamefont {M.-H.}\ \bibnamefont {Du}}, \bibinfo {author} {\bibfnamefont
  {D.~J.}\ \bibnamefont {Singh}}, \bibinfo {author} {\bibfnamefont
  {A.}~\bibnamefont {Podlesnyak}}, \bibinfo {author} {\bibfnamefont
  {G.}~\bibnamefont {Ehlers}}, \bibinfo {author} {\bibfnamefont {M.~D.}\
  \bibnamefont {Lumsden}}, \ and\ \bibinfo {author} {\bibfnamefont {B.~C.}\
  \bibnamefont {Sales}},\ }\href {\doibase 10.1038/nmat3035} {\bibfield
  {journal} {\bibinfo  {journal} {Nature Materials}\ }\textbf {\bibinfo
  {volume} {10}},\ \bibinfo {pages} {614} (\bibinfo {year} {2011})}\BibitemShut
  {NoStop}%
\bibitem [{\citenamefont {Xu}\ \emph {et~al.}(2012)\citenamefont {Xu},
  \citenamefont {Liu}, \citenamefont {Alidoust}, \citenamefont {Neupane},
  \citenamefont {Qian}, \citenamefont {Belopolski}, \citenamefont {Denlinger},
  \citenamefont {Wang}, \citenamefont {Lin}, \citenamefont {Wray},
  \citenamefont {Landolt}, \citenamefont {Slomski}, \citenamefont {Dil},
  \citenamefont {Marcinkova}, \citenamefont {Morosan}, \citenamefont {Gibson},
  \citenamefont {Sankar}, \citenamefont {Chou}, \citenamefont {Cava},
  \citenamefont {Bansil},\ and\ \citenamefont {Hasan}}]{Xu2012}%
  \BibitemOpen
  \bibfield  {author} {\bibinfo {author} {\bibfnamefont {S.~Y.}\ \bibnamefont
  {Xu}}, \bibinfo {author} {\bibfnamefont {C.}~\bibnamefont {Liu}}, \bibinfo
  {author} {\bibfnamefont {N.}~\bibnamefont {Alidoust}}, \bibinfo {author}
  {\bibfnamefont {M.}~\bibnamefont {Neupane}}, \bibinfo {author} {\bibfnamefont
  {D.}~\bibnamefont {Qian}}, \bibinfo {author} {\bibfnamefont {I.}~\bibnamefont
  {Belopolski}}, \bibinfo {author} {\bibfnamefont {J.~D.}\ \bibnamefont
  {Denlinger}}, \bibinfo {author} {\bibfnamefont {Y.~J.}\ \bibnamefont {Wang}},
  \bibinfo {author} {\bibfnamefont {H.}~\bibnamefont {Lin}}, \bibinfo {author}
  {\bibfnamefont {L.~A.}\ \bibnamefont {Wray}}, \bibinfo {author}
  {\bibfnamefont {G.}~\bibnamefont {Landolt}}, \bibinfo {author} {\bibfnamefont
  {B.}~\bibnamefont {Slomski}}, \bibinfo {author} {\bibfnamefont {J.~H.}\
  \bibnamefont {Dil}}, \bibinfo {author} {\bibfnamefont {A.}~\bibnamefont
  {Marcinkova}}, \bibinfo {author} {\bibfnamefont {E.}~\bibnamefont {Morosan}},
  \bibinfo {author} {\bibfnamefont {Q.}~\bibnamefont {Gibson}}, \bibinfo
  {author} {\bibfnamefont {R.}~\bibnamefont {Sankar}}, \bibinfo {author}
  {\bibfnamefont {F.~C.}\ \bibnamefont {Chou}}, \bibinfo {author}
  {\bibfnamefont {R.~J.}\ \bibnamefont {Cava}}, \bibinfo {author}
  {\bibfnamefont {A.}~\bibnamefont {Bansil}}, \ and\ \bibinfo {author}
  {\bibfnamefont {M.~Z.}\ \bibnamefont {Hasan}},\ }\href@noop {} {\bibfield
  {journal} {\bibinfo  {journal} {Nature Communications}\ }\textbf {\bibinfo
  {volume} {3}},\ \bibinfo {pages} {1192} (\bibinfo {year} {2012})}\BibitemShut
  {NoStop}%
\bibitem [{\citenamefont {Dziawa}\ \emph {et~al.}(2012)\citenamefont {Dziawa},
  \citenamefont {Kowalski}, \citenamefont {Dybko}, \citenamefont {Buczko},
  \citenamefont {Szczerbakow}, \citenamefont {Szot}, \citenamefont
  {{\L}usakowska}, \citenamefont {Balasubramanian}, \citenamefont {Wojek},
  \citenamefont {Berntsen}, \citenamefont {Tjernberg},\ and\ \citenamefont
  {Story}}]{Dziawa2012}%
  \BibitemOpen
  \bibfield  {author} {\bibinfo {author} {\bibfnamefont {P.}~\bibnamefont
  {Dziawa}}, \bibinfo {author} {\bibfnamefont {B.~J.}\ \bibnamefont
  {Kowalski}}, \bibinfo {author} {\bibfnamefont {K.}~\bibnamefont {Dybko}},
  \bibinfo {author} {\bibfnamefont {R.}~\bibnamefont {Buczko}}, \bibinfo
  {author} {\bibfnamefont {A.}~\bibnamefont {Szczerbakow}}, \bibinfo {author}
  {\bibfnamefont {M.}~\bibnamefont {Szot}}, \bibinfo {author} {\bibfnamefont
  {E.}~\bibnamefont {{\L}usakowska}}, \bibinfo {author} {\bibfnamefont
  {T.}~\bibnamefont {Balasubramanian}}, \bibinfo {author} {\bibfnamefont
  {B.~M.}\ \bibnamefont {Wojek}}, \bibinfo {author} {\bibfnamefont {M.~H.}\
  \bibnamefont {Berntsen}}, \bibinfo {author} {\bibfnamefont {O.}~\bibnamefont
  {Tjernberg}}, \ and\ \bibinfo {author} {\bibfnamefont {T.}~\bibnamefont
  {Story}},\ }\href@noop {} {\bibfield  {journal} {\bibinfo  {journal} {Nature
  Materials}\ }\textbf {\bibinfo {volume} {11}},\ \bibinfo {pages} {1023}
  (\bibinfo {year} {2012})}\BibitemShut {NoStop}%
\bibitem [{\citenamefont {Chernik}\ and\ \citenamefont
  {Lykov}(1981{\natexlab{a}})}]{Chernik1981}%
  \BibitemOpen
  \bibfield  {author} {\bibinfo {author} {\bibfnamefont {I.}~\bibnamefont
  {Chernik}}\ and\ \bibinfo {author} {\bibfnamefont {S.}~\bibnamefont
  {Lykov}},\ }\href@noop {} {\bibfield  {journal} {\bibinfo  {journal} {Soviet
  Physics, Solid State}\ }\textbf {\bibinfo {volume} {23}},\ \bibinfo {pages}
  {817} (\bibinfo {year} {1981}{\natexlab{a}})}\BibitemShut {NoStop}%
\bibitem [{\citenamefont {Matsushita}\ \emph {et~al.}(2006)\citenamefont
  {Matsushita}, \citenamefont {Wianecki}, \citenamefont {Sommer}, \citenamefont
  {Geballe},\ and\ \citenamefont {Fisher}}]{Matsushita2006}%
  \BibitemOpen
  \bibfield  {author} {\bibinfo {author} {\bibfnamefont {Y.}~\bibnamefont
  {Matsushita}}, \bibinfo {author} {\bibfnamefont {P.~A.}\ \bibnamefont
  {Wianecki}}, \bibinfo {author} {\bibfnamefont {A.~T.}\ \bibnamefont
  {Sommer}}, \bibinfo {author} {\bibfnamefont {T.~H.}\ \bibnamefont {Geballe}},
  \ and\ \bibinfo {author} {\bibfnamefont {I.~R.}\ \bibnamefont {Fisher}},\
  }\href {\doibase 10.1103/PhysRevB.74.134512} {\bibfield  {journal} {\bibinfo
  {journal} {Physical Review B}\ }\textbf {\bibinfo {volume} {74}},\ \bibinfo
  {pages} {134512} (\bibinfo {year} {2006})}\BibitemShut {NoStop}%
\bibitem [{\citenamefont {Giraldo-Gallo}\ \emph {et~al.}(2017)\citenamefont
  {Giraldo-Gallo}, \citenamefont {Walmsley}, \citenamefont {Sangiorgio},
  \citenamefont {Riggs}, \citenamefont {McDonald}, \citenamefont {Buchauer},
  \citenamefont {Fauque}, \citenamefont {Liu}, \citenamefont {Spaldin},
  \citenamefont {Kaminski}, \citenamefont {Behnia},\ and\ \citenamefont
  {Fisher}}]{Giraldo-Gallo2017}%
  \BibitemOpen
  \bibfield  {author} {\bibinfo {author} {\bibfnamefont {P.}~\bibnamefont
  {Giraldo-Gallo}}, \bibinfo {author} {\bibfnamefont {P.}~\bibnamefont
  {Walmsley}}, \bibinfo {author} {\bibfnamefont {B.}~\bibnamefont
  {Sangiorgio}}, \bibinfo {author} {\bibfnamefont {S.~C.}\ \bibnamefont
  {Riggs}}, \bibinfo {author} {\bibfnamefont {R.~D.}\ \bibnamefont {McDonald}},
  \bibinfo {author} {\bibfnamefont {L.}~\bibnamefont {Buchauer}}, \bibinfo
  {author} {\bibfnamefont {B.}~\bibnamefont {Fauque}}, \bibinfo {author}
  {\bibfnamefont {C.}~\bibnamefont {Liu}}, \bibinfo {author} {\bibfnamefont
  {N.~A.}\ \bibnamefont {Spaldin}}, \bibinfo {author} {\bibfnamefont
  {A.}~\bibnamefont {Kaminski}}, \bibinfo {author} {\bibfnamefont
  {K.}~\bibnamefont {Behnia}}, \ and\ \bibinfo {author} {\bibfnamefont {I.~R.}\
  \bibnamefont {Fisher}},\ }\href {http://arxiv.org/abs/1711.05723} {\
  (\bibinfo {year} {2017})},\ \Eprint {http://arxiv.org/abs/1711.05723}
  {arXiv:1711.05723} \BibitemShut {NoStop}%
\bibitem [{\citenamefont {Matsushita}\ \emph {et~al.}(2005)\citenamefont
  {Matsushita}, \citenamefont {Bluhm}, \citenamefont {Geballe},\ and\
  \citenamefont {Fisher}}]{Matsushita2005}%
  \BibitemOpen
  \bibfield  {author} {\bibinfo {author} {\bibfnamefont {Y.}~\bibnamefont
  {Matsushita}}, \bibinfo {author} {\bibfnamefont {H.}~\bibnamefont {Bluhm}},
  \bibinfo {author} {\bibfnamefont {T.~H.}\ \bibnamefont {Geballe}}, \ and\
  \bibinfo {author} {\bibfnamefont {I.~R.}\ \bibnamefont {Fisher}},\ }\href
  {\doibase 10.1103/PhysRevLett.94.157002} {\bibfield  {journal} {\bibinfo
  {journal} {Physical Review Letters}\ }\textbf {\bibinfo {volume} {94}},\
  \bibinfo {pages} {157002} (\bibinfo {year} {2005})},\ \Eprint
  {http://arxiv.org/abs/0409174} {0409174} \BibitemShut {NoStop}%
\bibitem [{\citenamefont {Singh}(2010)}]{Singh2010}%
  \BibitemOpen
  \bibfield  {author} {\bibinfo {author} {\bibfnamefont {D.~J.}\ \bibnamefont
  {Singh}},\ }\href {\doibase 10.1103/PhysRevB.81.195217} {\bibfield  {journal}
  {\bibinfo  {journal} {Physical Review B}\ }\textbf {\bibinfo {volume} {81}},\
  \bibinfo {pages} {195217} (\bibinfo {year} {2010})}\BibitemShut {NoStop}%
\bibitem [{\citenamefont {Giraldo-Gallo}\ \emph {et~al.}(2016)\citenamefont
  {Giraldo-Gallo}, \citenamefont {Sangiorgio}, \citenamefont {Walmsley},
  \citenamefont {Silverstein}, \citenamefont {Fechner}, \citenamefont {Riggs},
  \citenamefont {Geballe}, \citenamefont {Spaldin},\ and\ \citenamefont
  {Fisher}}]{Giraldo-Gallo2016}%
  \BibitemOpen
  \bibfield  {author} {\bibinfo {author} {\bibfnamefont {P.}~\bibnamefont
  {Giraldo-Gallo}}, \bibinfo {author} {\bibfnamefont {B.}~\bibnamefont
  {Sangiorgio}}, \bibinfo {author} {\bibfnamefont {P.}~\bibnamefont
  {Walmsley}}, \bibinfo {author} {\bibfnamefont {H.~J.}\ \bibnamefont
  {Silverstein}}, \bibinfo {author} {\bibfnamefont {M.}~\bibnamefont
  {Fechner}}, \bibinfo {author} {\bibfnamefont {S.~C.}\ \bibnamefont {Riggs}},
  \bibinfo {author} {\bibfnamefont {T.~H.}\ \bibnamefont {Geballe}}, \bibinfo
  {author} {\bibfnamefont {N.~A.}\ \bibnamefont {Spaldin}}, \ and\ \bibinfo
  {author} {\bibfnamefont {I.~R.}\ \bibnamefont {Fisher}},\ }\href {\doibase
  10.1103/PhysRevB.94.195141} {\bibfield  {journal} {\bibinfo  {journal}
  {Physical Review B}\ }\textbf {\bibinfo {volume} {94}},\ \bibinfo {pages}
  {195141} (\bibinfo {year} {2016})}\BibitemShut {NoStop}%
\bibitem [{\citenamefont {Ahmad}\ \emph {et~al.}(2006)\citenamefont {Ahmad},
  \citenamefont {Mahanti}, \citenamefont {Hoang},\ and\ \citenamefont
  {Kanatzidis}}]{Ahmad2006}%
  \BibitemOpen
  \bibfield  {author} {\bibinfo {author} {\bibfnamefont {S.}~\bibnamefont
  {Ahmad}}, \bibinfo {author} {\bibfnamefont {S.~D.}\ \bibnamefont {Mahanti}},
  \bibinfo {author} {\bibfnamefont {K.}~\bibnamefont {Hoang}}, \ and\ \bibinfo
  {author} {\bibfnamefont {M.~G.}\ \bibnamefont {Kanatzidis}},\ }\href
  {\doibase 10.1103/PhysRevB.74.155205} {\bibfield  {journal} {\bibinfo
  {journal} {Physical Review B}\ }\textbf {\bibinfo {volume} {74}},\ \bibinfo
  {pages} {155205} (\bibinfo {year} {2006})}\BibitemShut {NoStop}%
\bibitem [{\citenamefont {Xiong}\ \emph {et~al.}(2010)\citenamefont {Xiong},
  \citenamefont {Lee}, \citenamefont {Gupta}, \citenamefont {Wang},
  \citenamefont {Gnade},\ and\ \citenamefont {Cho}}]{Xiong2010}%
  \BibitemOpen
  \bibfield  {author} {\bibinfo {author} {\bibfnamefont {K.}~\bibnamefont
  {Xiong}}, \bibinfo {author} {\bibfnamefont {G.}~\bibnamefont {Lee}}, \bibinfo
  {author} {\bibfnamefont {R.~P.}\ \bibnamefont {Gupta}}, \bibinfo {author}
  {\bibfnamefont {W.}~\bibnamefont {Wang}}, \bibinfo {author} {\bibfnamefont
  {B.~E.}\ \bibnamefont {Gnade}}, \ and\ \bibinfo {author} {\bibfnamefont
  {K.}~\bibnamefont {Cho}},\ }\href {\doibase 10.1088/0022-3727/43/40/405403}
  {\bibfield  {journal} {\bibinfo  {journal} {Journal of Physics D: Applied
  Physics}\ }\textbf {\bibinfo {volume} {43}},\ \bibinfo {pages} {405403}
  (\bibinfo {year} {2010})}\BibitemShut {NoStop}%
\bibitem [{\citenamefont {Ravich}(1970)}]{Ravich1970}%
  \BibitemOpen
  \bibfield  {author} {\bibinfo {author} {\bibfnamefont {Y.}~\bibnamefont
  {Ravich}},\ }\href@noop {} {\emph {\bibinfo {title} {{Semiconducting Lead
  Chalcogenides}}}}\ (\bibinfo  {publisher} {Plenum Press},\ \bibinfo {year}
  {1970})\BibitemShut {NoStop}%
\bibitem [{\citenamefont {Nimtz}\ \emph {et~al.}(1983)\citenamefont {Nimtz},
  \citenamefont {Schlicht},\ and\ \citenamefont {Dornhaus}}]{Nimtz1983}%
  \BibitemOpen
  \bibfield  {author} {\bibinfo {author} {\bibfnamefont {G.}~\bibnamefont
  {Nimtz}}, \bibinfo {author} {\bibfnamefont {B.}~\bibnamefont {Schlicht}}, \
  and\ \bibinfo {author} {\bibfnamefont {R.}~\bibnamefont {Dornhaus}},\
  }\href@noop {} {\emph {\bibinfo {title} {{Narrow-Gap Semiconductors}}}}\
  (\bibinfo  {publisher} {Springer-Verlag},\ \bibinfo {year}
  {1983})\BibitemShut {NoStop}%
\bibitem [{\citenamefont {Allgaier}\ and\ \citenamefont
  {Houston}(1966)}]{Allgaier1966}%
  \BibitemOpen
  \bibfield  {author} {\bibinfo {author} {\bibfnamefont {R.~S.}\ \bibnamefont
  {Allgaier}}\ and\ \bibinfo {author} {\bibfnamefont {B.~B.}\ \bibnamefont
  {Houston}},\ }\href {\doibase 10.1063/1.1707831} {\bibfield  {journal}
  {\bibinfo  {journal} {Journal of Applied Physics}\ }\textbf {\bibinfo
  {volume} {37}},\ \bibinfo {pages} {302} (\bibinfo {year} {1966})}\BibitemShut
  {NoStop}%
\bibitem [{\citenamefont {Bilc}\ \emph {et~al.}(2006)\citenamefont {Bilc},
  \citenamefont {Mahanti},\ and\ \citenamefont {Kanatzidis}}]{Bilc2006}%
  \BibitemOpen
  \bibfield  {author} {\bibinfo {author} {\bibfnamefont {D.~I.}\ \bibnamefont
  {Bilc}}, \bibinfo {author} {\bibfnamefont {S.~D.}\ \bibnamefont {Mahanti}}, \
  and\ \bibinfo {author} {\bibfnamefont {M.~G.}\ \bibnamefont {Kanatzidis}},\
  }\href {\doibase 10.1103/PhysRevB.74.125202} {\bibfield  {journal} {\bibinfo
  {journal} {Physical Review B}\ }\textbf {\bibinfo {volume} {74}},\ \bibinfo
  {pages} {125202} (\bibinfo {year} {2006})}\BibitemShut {NoStop}%
\bibitem [{\citenamefont {Gibbs}\ \emph {et~al.}(2013)\citenamefont {Gibbs},
  \citenamefont {Kim}, \citenamefont {Wang}, \citenamefont {White},
  \citenamefont {Drymiotis}, \citenamefont {Kaviany},\ and\ \citenamefont
  {{Jeffrey Snyder}}}]{Gibbs2013}%
  \BibitemOpen
  \bibfield  {author} {\bibinfo {author} {\bibfnamefont {Z.~M.}\ \bibnamefont
  {Gibbs}}, \bibinfo {author} {\bibfnamefont {H.}~\bibnamefont {Kim}}, \bibinfo
  {author} {\bibfnamefont {H.}~\bibnamefont {Wang}}, \bibinfo {author}
  {\bibfnamefont {R.~L.}\ \bibnamefont {White}}, \bibinfo {author}
  {\bibfnamefont {F.}~\bibnamefont {Drymiotis}}, \bibinfo {author}
  {\bibfnamefont {M.}~\bibnamefont {Kaviany}}, \ and\ \bibinfo {author}
  {\bibfnamefont {G.}~\bibnamefont {{Jeffrey Snyder}}},\ }\href@noop {}
  {\bibfield  {journal} {\bibinfo  {journal} {Applied Physics Letters}\
  }\textbf {\bibinfo {volume} {103}},\ \bibinfo {pages} {262109} (\bibinfo
  {year} {2013})}\BibitemShut {NoStop}%
\bibitem [{\citenamefont {Zhao}\ \emph {et~al.}(2014)\citenamefont {Zhao},
  \citenamefont {Malliakas}, \citenamefont {Appathurai}, \citenamefont
  {Karlapati}, \citenamefont {Chung}, \citenamefont {Rosenkranz}, \citenamefont
  {Kanatzidis},\ and\ \citenamefont {Chatterjee}}]{Zhao2014}%
  \BibitemOpen
  \bibfield  {author} {\bibinfo {author} {\bibfnamefont {J.}~\bibnamefont
  {Zhao}}, \bibinfo {author} {\bibfnamefont {C.~D.}\ \bibnamefont {Malliakas}},
  \bibinfo {author} {\bibfnamefont {N.}~\bibnamefont {Appathurai}}, \bibinfo
  {author} {\bibfnamefont {V.}~\bibnamefont {Karlapati}}, \bibinfo {author}
  {\bibfnamefont {D.~Y.}\ \bibnamefont {Chung}}, \bibinfo {author}
  {\bibfnamefont {S.}~\bibnamefont {Rosenkranz}}, \bibinfo {author}
  {\bibfnamefont {M.~G.}\ \bibnamefont {Kanatzidis}}, \ and\ \bibinfo {author}
  {\bibfnamefont {U.}~\bibnamefont {Chatterjee}},\ }\href
  {http://arxiv.org/abs/1404.1807} {\  (\bibinfo {year} {2014})},\ \Eprint
  {http://arxiv.org/abs/1404.1807} {arXiv:1404.1807} \BibitemShut {NoStop}%
\bibitem [{\citenamefont {Jensen}\ \emph {et~al.}(1978)\citenamefont {Jensen},
  \citenamefont {Houston},\ and\ \citenamefont {Burke}}]{Jensen1978}%
  \BibitemOpen
  \bibfield  {author} {\bibinfo {author} {\bibfnamefont {J.~D.}\ \bibnamefont
  {Jensen}}, \bibinfo {author} {\bibfnamefont {B.}~\bibnamefont {Houston}}, \
  and\ \bibinfo {author} {\bibfnamefont {J.~R.}\ \bibnamefont {Burke}},\
  }\href@noop {} {\bibfield  {journal} {\bibinfo  {journal} {Physical Review
  B}\ }\textbf {\bibinfo {volume} {18}},\ \bibinfo {pages} {5567} (\bibinfo
  {year} {1978})}\BibitemShut {NoStop}%
\bibitem [{\citenamefont {Sitter}\ \emph {et~al.}(1977)\citenamefont {Sitter},
  \citenamefont {Lischka},\ and\ \citenamefont {Heinrich}}]{Sitter1977}%
  \BibitemOpen
  \bibfield  {author} {\bibinfo {author} {\bibfnamefont {H.}~\bibnamefont
  {Sitter}}, \bibinfo {author} {\bibfnamefont {K.}~\bibnamefont {Lischka}}, \
  and\ \bibinfo {author} {\bibfnamefont {H.}~\bibnamefont {Heinrich}},\
  }\href@noop {} {\bibfield  {journal} {\bibinfo  {journal} {Physical Review
  B}\ }\textbf {\bibinfo {volume} {16}},\ \bibinfo {pages} {680} (\bibinfo
  {year} {1977})}\BibitemShut {NoStop}%
\bibitem [{\citenamefont {Nakayama}\ \emph {et~al.}(2008)\citenamefont
  {Nakayama}, \citenamefont {Sato}, \citenamefont {Takahashi},\ and\
  \citenamefont {Murakami}}]{Nakayama2008}%
  \BibitemOpen
  \bibfield  {author} {\bibinfo {author} {\bibfnamefont {K.}~\bibnamefont
  {Nakayama}}, \bibinfo {author} {\bibfnamefont {T.}~\bibnamefont {Sato}},
  \bibinfo {author} {\bibfnamefont {T.}~\bibnamefont {Takahashi}}, \ and\
  \bibinfo {author} {\bibfnamefont {H.}~\bibnamefont {Murakami}},\ }\href
  {\doibase 10.1103/PhysRevLett.100.227004} {\bibfield  {journal} {\bibinfo
  {journal} {Physical Review Letters}\ }\textbf {\bibinfo {volume} {100}},\
  \bibinfo {pages} {227004} (\bibinfo {year} {2008})}\BibitemShut {NoStop}%
\bibitem [{\citenamefont {Yamini}\ \emph {et~al.}(2013)\citenamefont {Yamini},
  \citenamefont {Ikeda}, \citenamefont {Lalonde}, \citenamefont {Pei},
  \citenamefont {Dou},\ and\ \citenamefont {Snyder}}]{Yamini2013}%
  \BibitemOpen
  \bibfield  {author} {\bibinfo {author} {\bibfnamefont {S.~A.}\ \bibnamefont
  {Yamini}}, \bibinfo {author} {\bibfnamefont {T.}~\bibnamefont {Ikeda}},
  \bibinfo {author} {\bibfnamefont {A.}~\bibnamefont {Lalonde}}, \bibinfo
  {author} {\bibfnamefont {Y.}~\bibnamefont {Pei}}, \bibinfo {author}
  {\bibfnamefont {S.~X.}\ \bibnamefont {Dou}}, \ and\ \bibinfo {author}
  {\bibfnamefont {G.~J.}\ \bibnamefont {Snyder}},\ }\href {\doibase
  10.1039/c3ta11654a} {\bibfield  {journal} {\bibinfo  {journal} {Journal of
  Materials Chemistry A}\ }\textbf {\bibinfo {volume} {1}},\ \bibinfo {pages}
  {8725} (\bibinfo {year} {2013})}\BibitemShut {NoStop}%
\bibitem [{\citenamefont {{Van Degrift}}(1975)}]{VanDegrift1975}%
  \BibitemOpen
  \bibfield  {author} {\bibinfo {author} {\bibfnamefont {C.~T.}\ \bibnamefont
  {{Van Degrift}}},\ }\href {\doibase 10.1063/1.1134272} {\bibfield  {journal}
  {\bibinfo  {journal} {Review of Scientific Instruments}\ }\textbf {\bibinfo
  {volume} {46}},\ \bibinfo {pages} {599} (\bibinfo {year} {1975})}\BibitemShut
  {NoStop}%
\bibitem [{\citenamefont {Crocker}(1967)}]{Crocker1967}%
  \BibitemOpen
  \bibfield  {author} {\bibinfo {author} {\bibfnamefont {A.}~\bibnamefont
  {Crocker}},\ }\href {\doibase 10.1016/0022-3697(67)90167-9} {\bibfield
  {journal} {\bibinfo  {journal} {Journal of Physics and Chemistry of Solids}\
  }\textbf {\bibinfo {volume} {28}},\ \bibinfo {pages} {1903} (\bibinfo {year}
  {1967})}\BibitemShut {NoStop}%
\bibitem [{\citenamefont {Chernik}\ and\ \citenamefont
  {Lykov}(1981{\natexlab{b}})}]{Chernik1981b}%
  \BibitemOpen
  \bibfield  {author} {\bibinfo {author} {\bibfnamefont {I.}~\bibnamefont
  {Chernik}}\ and\ \bibinfo {author} {\bibfnamefont {S.}~\bibnamefont
  {Lykov}},\ }\href@noop {} {\bibfield  {journal} {\bibinfo  {journal} {Soviet
  Physics, Solid State}\ }\textbf {\bibinfo {volume} {23}},\ \bibinfo {pages}
  {1724} (\bibinfo {year} {1981}{\natexlab{b}})}\BibitemShut {NoStop}%
\bibitem [{\citenamefont {Kaidanov}\ and\ \citenamefont
  {Yu}(1985)}]{Kaidanov1985}%
  \BibitemOpen
  \bibfield  {author} {\bibinfo {author} {\bibfnamefont {V.~I.}\ \bibnamefont
  {Kaidanov}}\ and\ \bibinfo {author} {\bibfnamefont {I.~R.}\ \bibnamefont
  {Yu}},\ }\href {\doibase 10.1070/PU1985v028n01ABEH003632} {\bibfield
  {journal} {\bibinfo  {journal} {Soviet Physics Uspekhi}\ }\textbf {\bibinfo
  {volume} {28}},\ \bibinfo {pages} {31} (\bibinfo {year} {1985})}\BibitemShut
  {NoStop}%
\bibitem [{\citenamefont {Kaidanov}\ \emph {et~al.}(1989)\citenamefont
  {Kaidanov}, \citenamefont {Rykov},\ and\ \citenamefont
  {Rykova}}]{Kaidanov1989}%
  \BibitemOpen
  \bibfield  {author} {\bibinfo {author} {\bibfnamefont {V.}~\bibnamefont
  {Kaidanov}}, \bibinfo {author} {\bibfnamefont {S.}~\bibnamefont {Rykov}}, \
  and\ \bibinfo {author} {\bibfnamefont {M.}~\bibnamefont {Rykova}},\
  }\href@noop {} {\bibfield  {journal} {\bibinfo  {journal} {Soviet Physics,
  Solid State}\ }\textbf {\bibinfo {volume} {31}},\ \bibinfo {pages} {1316}
  (\bibinfo {year} {1989})}\BibitemShut {NoStop}%
\bibitem [{\citenamefont {Prokofieva}\ \emph {et~al.}(2009)\citenamefont
  {Prokofieva}, \citenamefont {Pshenay-Severin}, \citenamefont {Konstantinov},\
  and\ \citenamefont {Shabaldin}}]{Prokofieva2009}%
  \BibitemOpen
  \bibfield  {author} {\bibinfo {author} {\bibfnamefont {L.~V.}\ \bibnamefont
  {Prokofieva}}, \bibinfo {author} {\bibfnamefont {D.~A.}\ \bibnamefont
  {Pshenay-Severin}}, \bibinfo {author} {\bibfnamefont {P.~P.}\ \bibnamefont
  {Konstantinov}}, \ and\ \bibinfo {author} {\bibfnamefont {A.~A.}\
  \bibnamefont {Shabaldin}},\ }\href {\doibase 10.1134/S1063782609090097}
  {\bibfield  {journal} {\bibinfo  {journal} {Semiconductors}\ }\textbf
  {\bibinfo {volume} {43}},\ \bibinfo {pages} {1155} (\bibinfo {year}
  {2009})}\BibitemShut {NoStop}%
\bibitem [{Note1()}]{Note1}%
  \BibitemOpen
  \bibinfo {note} {The low-field {H}all coefficient modelled for two
  inequivalent hole-pockets is written as ${R}_{H} = {\begingroup 1\endgroup
  \over ec}{\begingroup A_{p_1}b^2_b p_{1}+A_{p_2}p_2\endgroup \over (b_p p_1
  +p_2)^2}$ where $p_i$ and ${A}_{p_i}$ are the hole density and {H}all factor
  of pocket $i$ respectively and the ratio of the mobilities of the pockets is
  $b_p={\begingroup \mu _{p_1}\endgroup \over \mu _{p_2}}$. The limiting case
  $\mu _{p_1}\gg \mu _{p_2}$ reduces to ${R}_{H}={\begingroup A_{p_1}\endgroup
  \over ecp_1}$, which is larger than the high-field value, ${R_H}({B} \to
  \infty )={\begingroup 1\endgroup \over ec(p_1+p_2)}$ provided ${A}_{p_i}$ is
  sufficiently close to unity. {R}emembering that we have defined $p_{H}({B}\to
  0)={\begingroup 1\endgroup \over R_H(B \to 0)e}$, it is therefore the case
  that $p_{H}({B}\to 0) < p$ for the two-band model in the absence of large
  anisotropies.}\BibitemShut {Stop}%
\bibitem [{\citenamefont {Hurd}(1972)}]{Hurd1972}%
  \BibitemOpen
  \bibfield  {author} {\bibinfo {author} {\bibfnamefont {C.~M.}\ \bibnamefont
  {Hurd}},\ }\href@noop {} {\emph {\bibinfo {title} {The Hall effect in metals
  and alloys}}}\ (\bibinfo  {publisher} {Plenum Press},\ \bibinfo {year}
  {1972})\BibitemShut {NoStop}%
\bibitem [{\citenamefont {Ong}(1991)}]{Ong1991}%
  \BibitemOpen
  \bibfield  {author} {\bibinfo {author} {\bibfnamefont {N.~P.}\ \bibnamefont
  {Ong}},\ }\href {\doibase 10.1103/PhysRevB.43.193} {\bibfield  {journal}
  {\bibinfo  {journal} {Physical Review B}\ }\textbf {\bibinfo {volume} {43}},\
  \bibinfo {pages} {193} (\bibinfo {year} {1991})}\BibitemShut {NoStop}%
\bibitem [{\citenamefont {Airapetyants}\ \emph {et~al.}(1966)\citenamefont
  {Airapetyants}, \citenamefont {Vinogradova}, \citenamefont {Dubrovskaya},
  \citenamefont {Kolomoets},\ and\ \citenamefont {Rudnik}}]{Airapetyants1966}%
  \BibitemOpen
  \bibfield  {author} {\bibinfo {author} {\bibfnamefont {S.~V.}\ \bibnamefont
  {Airapetyants}}, \bibinfo {author} {\bibfnamefont {M.~N.}\ \bibnamefont
  {Vinogradova}}, \bibinfo {author} {\bibfnamefont {I.~N.}\ \bibnamefont
  {Dubrovskaya}}, \bibinfo {author} {\bibfnamefont {N.~V.}\ \bibnamefont
  {Kolomoets}}, \ and\ \bibinfo {author} {\bibfnamefont {I.~M.}\ \bibnamefont
  {Rudnik}},\ }\href@noop {} {\bibfield  {journal} {\bibinfo  {journal} {Soviet
  Physics: Solid State}\ }\textbf {\bibinfo {volume} {8}},\ \bibinfo {pages}
  {1069} (\bibinfo {year} {1966})}\BibitemShut {NoStop}%
\bibitem [{\citenamefont {Kokhlov}(2003)}]{Kokhlov2003}%
  \BibitemOpen
  \bibfield  {author} {\bibinfo {author} {\bibfnamefont {D.}~\bibnamefont
  {Kokhlov}},\ }\href@noop {} {\emph {\bibinfo {title} {Lead Chalcogenides:
  Physics and Applications}}}\ (\bibinfo  {publisher} {Taylor and Francis},\
  \bibinfo {year} {2003})\BibitemShut {NoStop}%
\bibitem [{\citenamefont {Nemov}\ and\ \citenamefont
  {Ravich}(1998)}]{Nemov1998}%
  \BibitemOpen
  \bibfield  {author} {\bibinfo {author} {\bibfnamefont {S.}~\bibnamefont
  {Nemov}}\ and\ \bibinfo {author} {\bibfnamefont {Y.}~\bibnamefont {Ravich}},\
  }\href {\doibase 10.3367/UFNr.0168.199808a.0817} {\bibfield  {journal}
  {\bibinfo  {journal} {Uspekhi Fizicheskih Nauk}\ }\textbf {\bibinfo {volume}
  {168}},\ \bibinfo {pages} {817} (\bibinfo {year} {1998})}\BibitemShut
  {NoStop}%
\bibitem [{\citenamefont {Walmsley}\ \emph {et~al.}(2018)\citenamefont
  {Walmsley}, \citenamefont {Liu}, \citenamefont {Palczewski}, \citenamefont
  {Giraldo-Gallo}, \citenamefont {Olson}, \citenamefont {Fisher},\ and\
  \citenamefont {Kaminski}}]{Walmsley2018}%
  \BibitemOpen
  \bibfield  {author} {\bibinfo {author} {\bibfnamefont {P.}~\bibnamefont
  {Walmsley}}, \bibinfo {author} {\bibfnamefont {C.}~\bibnamefont {Liu}},
  \bibinfo {author} {\bibfnamefont {A.~D.}\ \bibnamefont {Palczewski}},
  \bibinfo {author} {\bibfnamefont {P.}~\bibnamefont {Giraldo-Gallo}}, \bibinfo
  {author} {\bibfnamefont {C.~G.}\ \bibnamefont {Olson}}, \bibinfo {author}
  {\bibfnamefont {I.~R.}\ \bibnamefont {Fisher}}, \ and\ \bibinfo {author}
  {\bibfnamefont {A.}~\bibnamefont {Kaminski}},\ }\href@noop {} {\  (\bibinfo
  {year} {2018})},\ \Eprint {http://arxiv.org/abs/1808.07213}
  {arXiv:1808.07213} \BibitemShut {NoStop}%
\end{thebibliography}
%

\end{document}